\documentclass[twocolumn,10pt]{asme2ej}

\usepackage{graphicx} 
\usepackage{hyperref}   
\hypersetup{
	colorlinks=true,
	linkcolor=blue,
	citecolor=blue,
	urlcolor=blue,
}
\usepackage[square,numbers]{natbib}
\usepackage{subfigure}
\usepackage{graphicx} 
\usepackage[noend]{algpseudocode}
\usepackage{algorithmicx,algorithm}
\usepackage{multirow}
\usepackage{xcolor}
\usepackage{amsmath}
\usepackage{enumitem}
\usepackage{booktabs}

\title{A Token-FCM based risk assessment method for complex engineering designs}

\author{Guan Wang
    \affiliation{
       \\ School of Information and Electrical Engineering\\ Hangzhou City University\\ Hangzhou,China\\
        Email:wangguan@hzcu.edu.cn
    }	
}

\author{Yimin Feng
    \affiliation{
       \\ School of Electronic Information and AI\\ Shaanxi University of Science and Technology\\ Xi'an,China\\
        Email:yiminfengmy@zju.edu.cn
    }	
}
\author{Rongbin Guo
    \affiliation{
       \\ School of Information and Electrical Engineering\\ Hangzhou City University\\ Hangzhou,China\\
        Email:rbguoisee@zju.edu.cn
    }	
}

\author{Yusheng Liu\thanks{ysliu@cad.zju.edu.}
    \affiliation{   
    \\ State Key Laboratory of CAD$\&$CG\\ Zhejiang University\\ Hangzhou,China\\
        Email:ysliu@cad.zju.edu.cn
    }
}

\author{Qiang Zou\thanks{qiangzou@cad.zju.edu.cn} 
    \affiliation{  \\ State Key Laboratory of CAD$\&$CG\\ Zhejiang University\\ Hangzhou,China\\
        Email:qiangzou@cad.zju.edu.cn
    }
}

\begin{document}

\maketitle    

\begin{abstract}
{
\it Engineering design risks could cause unaffordable losses, and thus risk assessment plays a critical role in engineering design. On the other hand, the high complexity of modern engineering designs makes it difficult to assess risks effectively and accurately due to the complex two-way, dynamic causal-effect risk relations in engineering designs. To address this problem, this paper proposes a new risk assessment method called token fuzzy cognitive map (Token-FCM). Its basic idea is to model the two-way causal-risk relations with the FCM method, and then augment FCM with a token mechanism to model the dynamics in causal-effect risk relations. Furthermore, the fuzzy sets and the group decision-making method are introduced to initialize the Token-FCM method so that comprehensive and accurate risk assessments can be attained. The effectiveness of the proposed method has been demonstrated by a real example of engine design for a horizontal directional drilling machine.
}
\end{abstract}


\section{Introduction}

Engineering designs have become an indispensable part of our lives, but all designs can potentially fail in operation, which then causes harm to humans and the environment. For this reason, risk assessment is often used in advance to help designers analyze and reduce risks before engineering products are implemented~\cite{10.1115/1.2901055}.

Fault tree analysis (FTA) and failure mode and effect analysis (FMEA) are the two dominant methods of risk assessment. They can model static causal-effect risk relations between design elements that operate in binary states but have limited applicability to dynamic scenarios~\cite{abdo2016uncertainty}. For such designs, there are time delays in the causal-effect relations of design risks, and design elements go beyond just static binary states (faulty and nonfaulty), but towards time-dependent multiple states.

\begin{figure*}[t]
    \centering
    \includegraphics[width=0.75\textwidth]{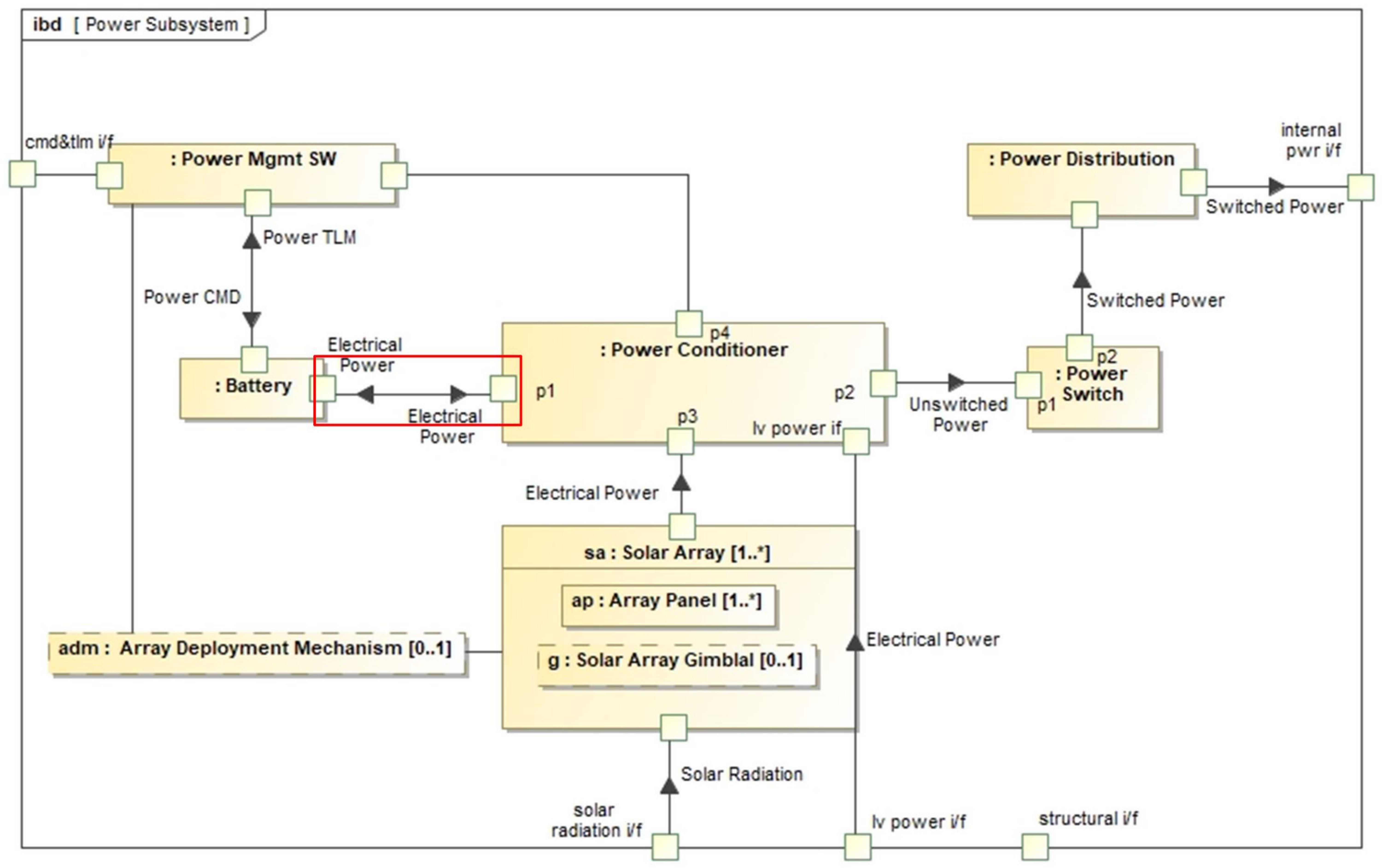}
    \caption{Process flow diagram of the emergency core cooling system.}
    \label{fig:diagram-of-cooling-system}
\end{figure*}

Typical examples of dynamic causal-effect risk relations are those requiring continuous operation, e.g., nuclear power plants. Figure~\ref{fig:diagram-of-cooling-system} shows the process flow of a nuclear power plant's emergency cooling system. When the In-containment Refueling Water Storage Tank (IRWST) leaks, the two Heater Changers (HX1, HX2) can still work normally for a while, but after that, the heat exchanger will fail due to lack of overheated cooling water. The corresponding fault point is indicated by the red rectangle in the figure.

\begin{figure*}[t]
    \centering
    \includegraphics[width=0.75\textwidth]{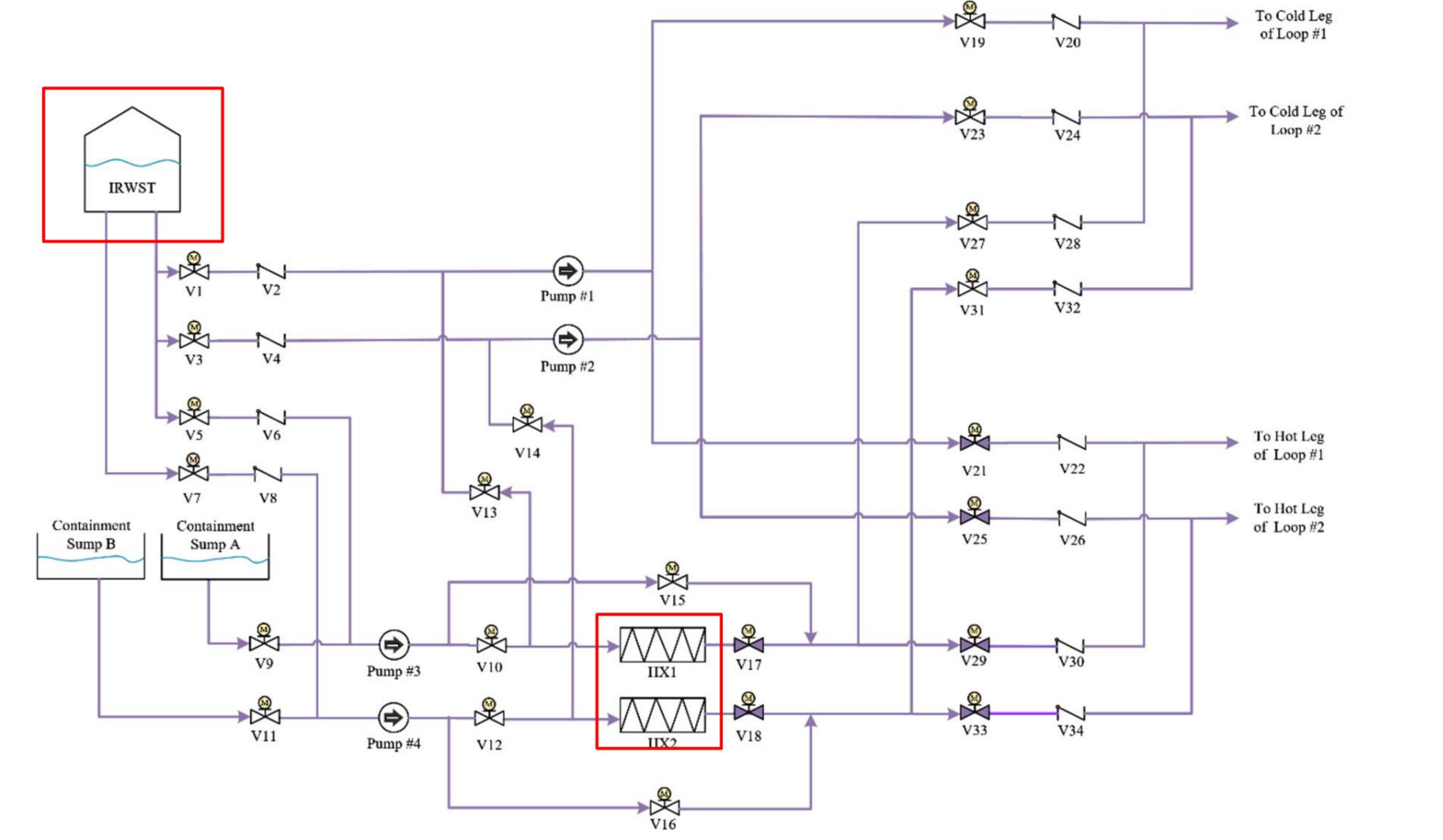}
    \caption{Internal block diagram of spacecraft power subsystem using the SysML.}
    \label{fig:diagram-of-power-subsystem}
\end{figure*}

To handle dynamics in causal-effect risk relations, several improvement methods, such as Petri nets~\cite{choi1994petri} and Bayesian networks~\cite{neil2012availability}, have been introduced. However, they consistently adopt tree structures to model design risks and consequently, can only deal with one-way causal-effect risk relations. Today, engineering designs are becoming increasingly complex and two-way causal-effect risk relations are increasingly seen~\cite{10.1115/1.4005230}. For such cases, faults at both ends of the interface of a system's components and/or parts will affect both sides~\citep{nair2020generalised}. One typical example is the spacecraft.  Figure~\ref{fig:diagram-of-power-subsystem} shows the internal structure relationship of the power supply subsystem of a spacecraft, where the battery failure (marked by the red rectangle) will affect the power conditioner and power conditioner failure will also cause damage to the battery.

Unfortunately, current risk assessment methods have limited applicability to engineering designs with two-way dynamic causal-effect risk relations. To this end, this paper proposes a new risk assessment method based on the fuzzy cognitive map (FCM)~\cite{kosko1986fuzzy}. FCM is a notion from the field of fuzzy logic theory, and it models the directed relation and the strength between two elements (e.g., states and events) by using weighted digraphs. Clearly, the FCM's graph structure is well-suited for modeling two-way causal-effect risk relations of engineering designs. Nevertheless, applying FCM to risk assessment is not straightforward for two reasons:
\begin{enumerate}
    \item FCM, in its original form, cannot simulate the dynamics of design risks~\citep{papageorgiou2012review}.    
    \item FCM's effectiveness relies heavily on node initialization, which is often done by expert scoring or the like~\citep{felix2019review}. This works for simple problems but becomes unsatisfactory when applied to engineering designs that involve complex risk sources and require comprehensive consideration of fault consequences, fault detection difficulties, fault occurring frequencies, etc.
\end{enumerate}

To solve these challenges, this paper proposes to augment FCM with a token mechanism to simulate and analyze time-dependent causal-effect risk relations (Section~\ref{sec:token-fcm}). This method is to be called Token-FCM hereafter. This paper also provides a comprehensive initialization method for Token-FCM, based on the combination of fuzzy sets and group decision-making (Section~\ref{sec:FCM-initialization}). FCM, the token mechanism, and the new initialization method work together to provide an effective risk assessment method for engineering designs with two-way and dynamic causal-effect risk relations, as confirmed by the engine design case in Section~\ref{sec:results}.

The remainder of this paper is as follows. Section~\ref{sec:relatework} reviews existing research studies. Section~\ref{sec:token-fcm} introduces the Token-FCM method, and Section~\ref{sec:overallmethod} presents the design risk assessment method based on Token-FCM. Section~\ref{sec:results} demonstrates the effectiveness of the proposed method using an example of engine design. Section~\ref{sec:conclusion} concludes the entire article.

\section{Related work}
\label{sec:relatework}
\subsection{Risk assessment methods}
There are two main risk assessment methods in the literature: FTA and FMEA. Their method details, advantages, and limitations have been documented in the literature~\cite{RUIJTERS201529,HUANG2020106885}. The most notable limitation is that these methods can only handle static systems. To overcome this limitation, several improvement methods such as Markov chain, Monte Carlo simulation, Bayesian network, and Petri net have been introduced, as detailed below.

The fundamental idea of the Markov chain method is to use the transition matrix to model risks in a probabilistic way~\cite{chandra1983machine,saydam2013time}. Such methods can provide a quantification of design risks, but they work only for systems whose risks follow an exponential distribution pattern. In addition, it suffers from the issue of state-space explosion. 

The Monte Carlo simulation method quantitatively assesses risks through random sampling and numerical simulation~\cite{laperriere2001monte,10.1115/1.4034219}. This method can be used to analyze various state spaces, and an approximate solution to the system can be obtained through sampling simulation. However, a great deal of numerical simulations must be performed to obtain an analysis with high fidelity, leading to high computational costs.

The Bayesian network method models a system as a network whose nodes, edges (representing causal-effect relations), and edge weights (representing relation strengths) are constructed using prior knowledge~\cite{khakzad2013quantitative,10.1115/1.4032399}. All risk analysis and reliability assessments are conducted by this knowledge network. The main limitations of Bayesian networks include: (1) the weak system semantics description ability~\cite{WEISS20051127}; (2) the difficulty in ensuring the consistency of the model; and (3) the large number of parameters~\cite{kabir2019applications}.

The Petri net method has clear semantics and rigorous mathematical expressions, which can help to create a coherent model of the behavior of the system~\cite{wu2015extended}. Petri net is constructed by identifying the state and transition of system behavior, specifying the transition trigger rate, and then evaluating system reliability. Despite these advantages, Petri net-based approaches often suffer from the problem of state-space explosion, which limits their application in the analysis of complex engineering systems~\cite{10.1115/1.3125203}.

The above methods have also been combined with each other to achieve hybrid versions. For example, the Markov chain and the Monte Carlo method have been combined to form Markov chain Monte Carlo approaches~\cite{kelly2009bayesian}, promoting the use of probabilistic reasoning in conventional statistical risk assessment methods. Mura and Bondavalli~\cite{mura2001markov} combined the Markov chain and the Petri net to model dynamic behavior and proposed an analysis technique having improved computational efficiency. Bai et al.~\cite{bai2005software} proposed a new Bayesian network method called the Markov Bayesian network, which constructs Bayesian networks through fault data obtained from historical data and expert knowledge. Cadini and Gioletta~\cite{cadini2016bayesian} introduced Bayesian methods into the Monte Carlo method in order to solve the deficiency of the Monte Carlo sampling method in estimating the probability of failure of the system under small values.

Nevertheless, numerous difficulties persist. For instance, techniques like FTA, Bayesian networks, and Petri nets evaluate design risks using tree structures, limiting them to unidirectional causal-effect risk associations. Today's engineering designs increasingly exhibit bidirectional causal-effect risk associations~\cite{KIM2012947}, as highlighted in the Introduction section. Likewise, existing approaches struggle with dynamic causal-effect risk relationships, such as those involving time delays. Consequently, there is a demand for further advancement of risk assessment methodologies to enhance their effectiveness for complex bidirectional dynamic causal-effect risk relationships in engineering designs. Moreover, traditional methods are fraught with issues like state space explosion and an abundance of parameters, which remain largely unresolved.

\subsection{The FCM method}
\begin{figure}[t]
    \centering
    \includegraphics[width=0.35\textwidth]{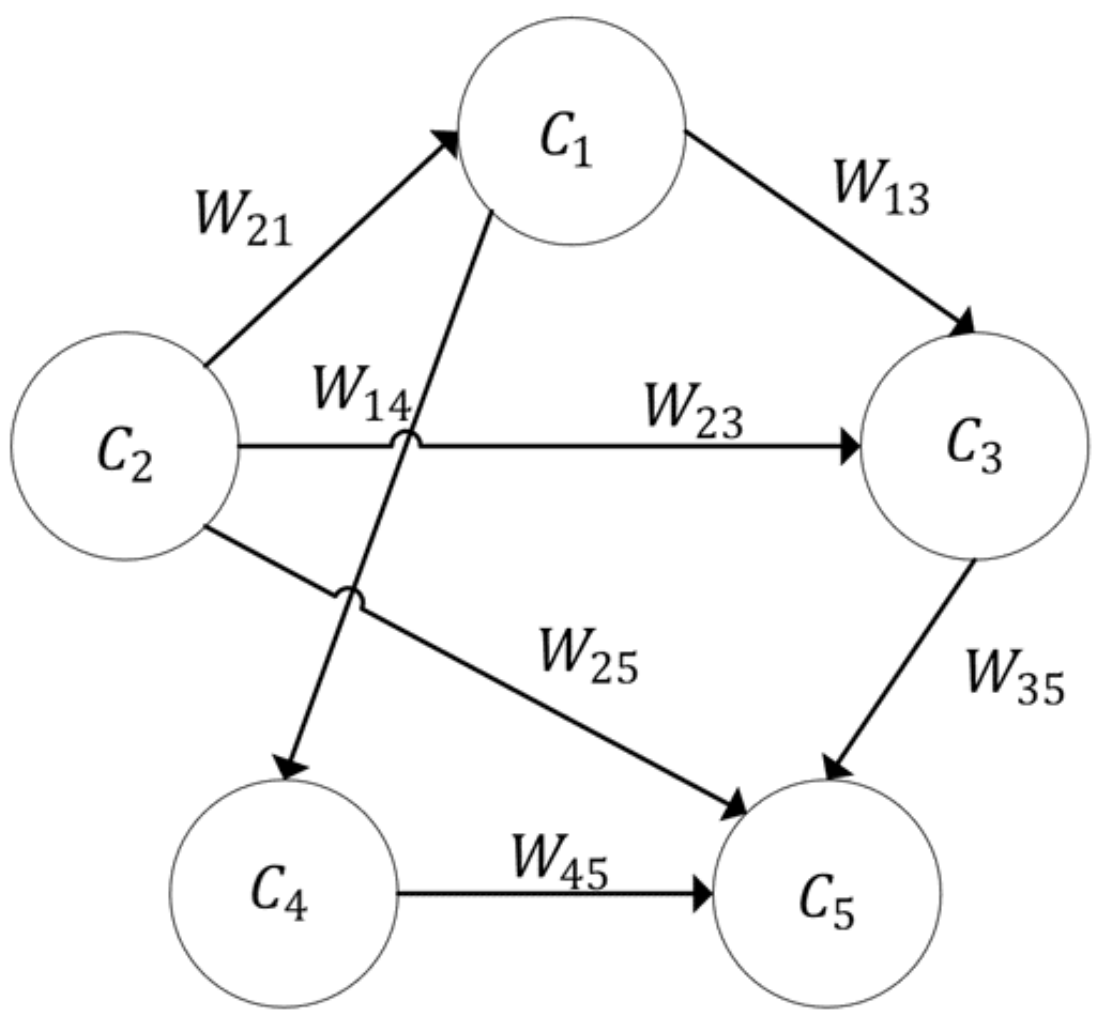}
    \caption{An example of FCM with five nodes and seven arcs.}
    \label{fig:FCM-diagram}
\end{figure}

As the proposed method in this work is built upon FCM, this subsection will provide a brief introduction to this method and review its published application to risk assessment. FCM is a soft computing technology introduced by Kosko~\cite{kosko1986fuzzy}. It combines neural networks with fuzzy logic to model the relation and degree of influence between concepts through feedback. An FCM is made up of nodes and weighted arcs (see Figure~\ref{fig:FCM-diagram}). In the context of complex engineering systems, nodes $C_i$ represent critical components, operational states, or design risks that could affect system performance. The selection of nodes depends on the architecture of the system, the hazards involved, and the key risk factors that need to be analyzed. Node values are initialized based on expert judgment or historical data, representing the risk or likelihood of failure for each component. For example, a high risk value may correspond to a critical design risk, while a low value indicates less risk.

The weighted arc $W_{ij}$ represents the extent of dependency or influence between the FCM nodes $C_i$ and $C_j$. The weight assigned to this arc indicates the degree of this relationship, which can be established via expert assessment, statistical analysis, or system modeling. Time delays model the time lag between cause and effect and can vary depending on the system's dynamic characteristics.

To enable comparison among different system components, both the node value and the arc weight are dimensionless. These metrics are calculated from standardized risk indices and causal influences, guaranteeing that the overall risk evaluation is also dimensionless. A scaling factor can be applied to consider thresholds specific to particular systems.

The inference mechanism of FCMs is similar to that of neural networks and is based on simple mathematical operations on the initial node values and weights. Specifically, in each iteration, the node value $C_i$ is recalculated, and the update process is~\cite{papageorgiou2013fuzzy}:

\begin{equation}
    C_i^{k+1}=f(C_i^k+\sum_{j=1,j\neq i}^{n}W_{ij}C_j^k)
\label{eq1}
\end{equation}
where $W_{ij}$ denote edge weight between nodes $i$ and $j$, $C_i^{k}$ represents $i$-th node's value at k-th iteration, $f$ is a threshold function bounding the node value $C_i$ into interval $[0,1]$. The iteration repeats until one of the following states is reached~\cite{papageorgiou2013fuzzy}: 

\begin{enumerate}
  \item  A maximum iteration number is reached.     
  \item FCM has transformed to a stable state. 
  \item Chaotic behavior has appeared.
\end{enumerate}

FCM is recognized as an effective and succinct tool for system analysis, offering benefits such as having a reduced number of system states and ease of computation, and it has been extensively applied in risk evaluation. Han and Deng~\cite{han2018hybrid} applied FCM to identify critical success factors for high-risk emergency systems. Lopez and Salmeron~\cite{lopez2014dynamic} used FCM to assess and manage risks in enterprise resource planning (ERP) systems, helping managers model complex objects and manage risks of ERP systems in a more effective manner. Lazzerini and Mkrtchyan~\cite{lazzerini2011analyzing} applied FCM to software project management, analyzing the relation between risks. Dabbagh and Yousefi~\cite{dabbagh2019hybrid} used FCM to identify factory occupational health and safety risks to minimize the negative consequences of these risks.

However, emergency systems, ERP systems, and software project management systems are quite different from engineering designs; in particular, they do not typically exhibit dynamic causal-risk relations. For this reason, FCM has not been used in the field of engineering design, to the best of the authors' knowledge. To overcome this limitation, we propose an enhanced version of FCM in the next section, which enhances FCM by incorporating a token mechanism in order to effectively model and simulate the dynamics (i.e., time delays) in causal-effect risk relations. Furthermore, a new FCM node initialization method technique is proposed that focuses on engineering design characteristics.

\section{The proposed Token-FCM method}
\label{sec:token-fcm}
\subsection{Overview of Token-FCM}
For the original FCM, all nodes are updated simultaneously. In contrast, the problem considered in this work has a strict chronological trigger sequence when updating nodes due to time delays in causal-effect risk relations. This gap is to be solved through Token-FCM which controls the chronological trigger sequence of node value updates through the intelligent behavior of tokens.

\begin{figure}[t]
    \centering
    \subfigure[]{%
    \resizebox*{6cm}{!}{\includegraphics{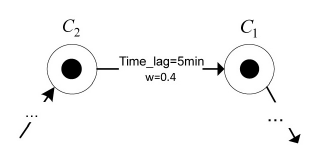}}}\hspace{5pt}
    \subfigure[]{%
    \resizebox*{6cm}{!}{\includegraphics{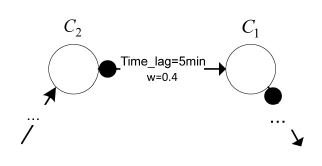}}}\hspace{5pt}
    \caption{Two states of the nodes: (a) activated and (b) inactivated. The black dots represent tokens, and the same for the following figures.}
    \label{fig:token-fcm}
\end{figure}

The structure of Token-FCM is depicted in Figure~\ref{fig:token-fcm}. It contains nodes, arcs, and tokens. The main feature of Token-FCM is that a node only updates its node value when it has a token and remains unchanged at other times. As such, each Token-FCM node has two modes when updating: activated and inactivated. By controlling the time delay, a node sends its token to another (neighboring) node, and all nodes can be updated in a predefined order and with desired time delays.

The high-level operation of Token-FCM is as follows. In the beginning, each node has one and only one token. Then, each token leaves its hosting node and moves to another node along the connecting arc (as shown in Figure~\ref{fig:token-fcm}b). Each arc is assigned a specific time delay and a token needs to wait for this time delay to reach the target node. Once reached, the token's attributes are updated and the target node is activated for value update. This process repeats until all nodes reach stable states.

\subsection{Token behavior design}
\label{sec:token-behavior-design}
The token-FCM contains various types of causal-effect relations, and tokens also exhibit different behaviors under different relations, so they need to be defined individually. We classify all causal-effect relations into three categories: (1) one-to-one; (2) one-to-many; and (3) many-to-one. Please note that the many-to-many case is a combination of many-to-one and one-to-many, and we omit the discussion of it.

\subsubsection{One-to-one causal-effect relations}
\begin{figure}[t]
    \centering
    \subfigure[]{%
    \resizebox*{5cm}{!}{\includegraphics{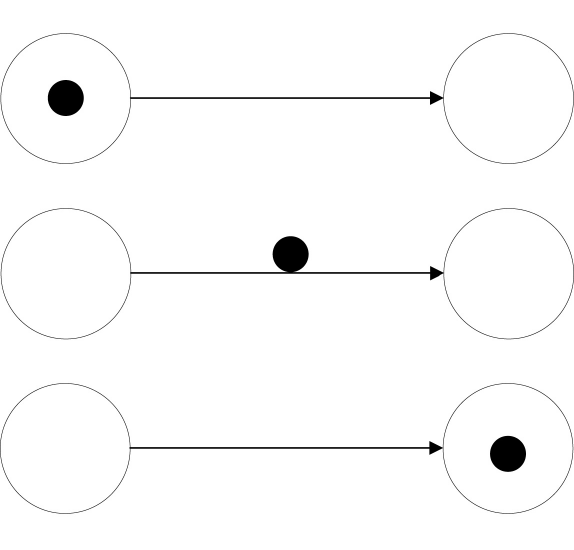}}}\hspace{5pt}
    \subfigure[]{%
    \resizebox*{5cm}{!}{\includegraphics{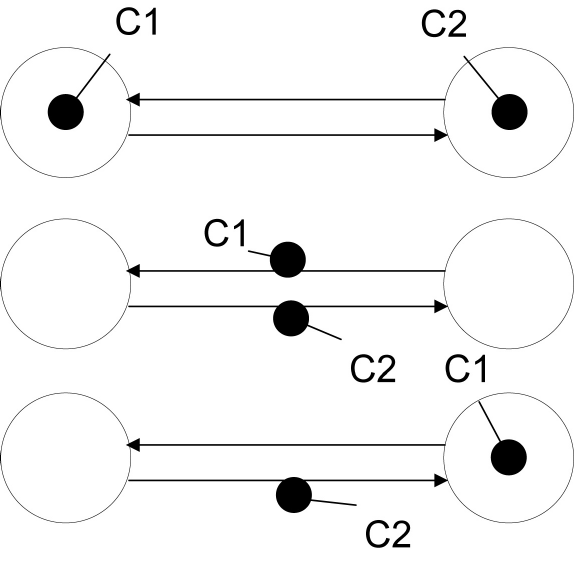}}}\hspace{5pt}
    \caption{Tokens behavior under one-to-one causal-effect relation: (a) One-way one-to-one causal-effect relation; (b) Two-way one-to-one causal-effect relation.}
    \label{fig:tokens-one-to-one-behavior}
\end{figure}

In this case, the token will reach the target node after a certain time delay along the arc. There are two relation types: one-way and two-way. Figure~\ref{fig:tokens-one-to-one-behavior}a shows a one-way one-to-one causal-effect relation. The token of the left node reaches the right node after a certain time delay along the arc and activates the right node. Figure~\ref{fig:tokens-one-to-one-behavior}b shows a two-way one-to-one causal-effect relation. As the time delays here have directions and could be different, a node's token may arrive at the other node asynchronously (the bottom example at Figure~\ref{fig:tokens-one-to-one-behavior}b).

\subsubsection{One-to-many causal-effect relation}
\begin{figure}[t]
    \centering
    \subfigure[]{%
    \resizebox*{3cm}{!}{\includegraphics{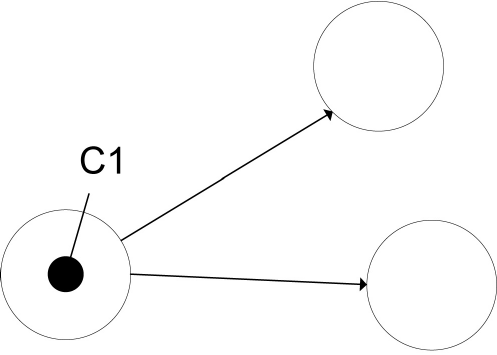}}}\hspace{5pt}
    \subfigure[]{%
    \resizebox*{3cm}{!}{\includegraphics{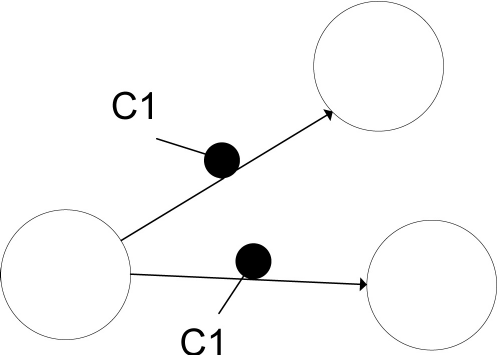}}}\hspace{5pt}
    \subfigure[]{%
    \resizebox*{3cm}{!}{\includegraphics{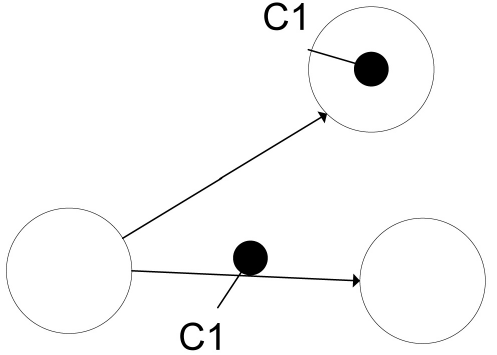}}}\hspace{5pt}
    \subfigure[]{%
    \resizebox*{3cm}{!}{\includegraphics{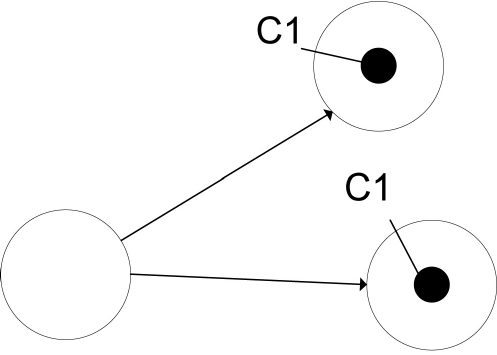}}}\hspace{5pt}
    \caption{Tokens behavior under one-to-many causal-effect relation.}
    \label{fig:tokens-one-to-many-behavior}
\end{figure}

If the risk of one component causes problems in multiple other components, a one-to-many causal-effect relation is formed. Using token-FCM, the token at the starting node is duplicated for all adjacent nodes in the beginning, and then all these tokens move away from the original node at the same time (Figure~\ref{fig:tokens-one-to-many-behavior}b). Their arrival at individual target nodes is controlled by the time delays in the arcs. Therefore, they may or may not arrive at the target nodes at the same time (Figures~\ref{fig:tokens-one-to-many-behavior}c and~\ref{fig:tokens-one-to-many-behavior}d).

\subsubsection{Many-to-one causal-effect relation}
\begin{figure}[htb]
    \centering
    \subfigure[]{%
    \resizebox*{3.8cm}{!}{\includegraphics{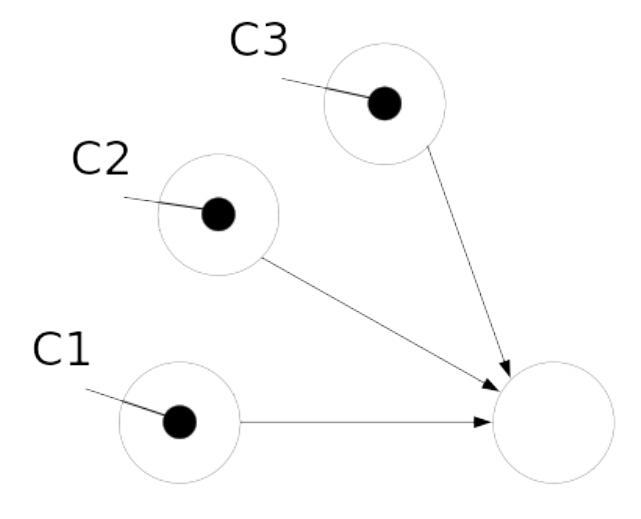}}}\hspace{5pt}
     \subfigure[]{%
     \resizebox*{3.8cm}{!}
     {\includegraphics{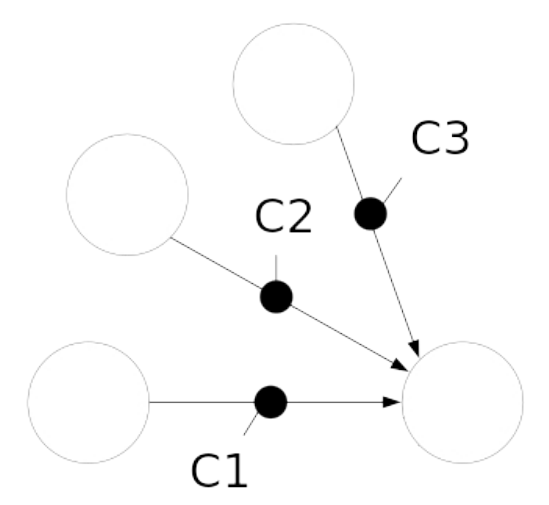}}}\hspace{5pt}
    \subfigure[]{%
    \resizebox*{3.8cm}{!}{\includegraphics{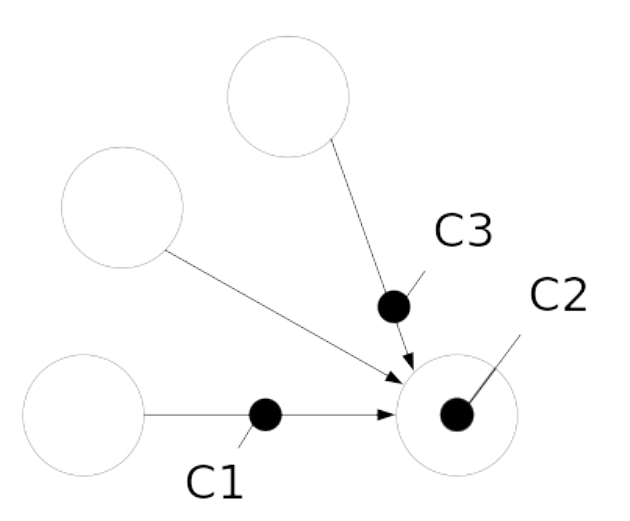}}}\hspace{5pt}
    \caption{Tokens behavior under many-to-one causal-effect relation.}
    \label{fig:tokens-many-to-one-behavior}
\end{figure}

One component may also be affected by the risk of multiple components, and this implies a many-to-one causal-effect relation. In this case, again, the arrival of all tokens at the target node is controlled by the time delay in the arcs, which means that each token has its own arrival time (Figure~\ref{fig:tokens-many-to-one-behavior}b). It should be noted that if the tokens arrive at different times, the node will be activated multiple times, thus performing multiple value updates.

\subsection{Token specification design}
The token in Token-FCM is responsible for processing time delay and activating node value updates. To do so, each token is designed to carry five kinds of information: 
\begin{enumerate}\setlength{\itemsep}{1pt}\setlength{\parskip}{0pt}\setlength{\parsep}{0pt}
    \item $Token\_ID$: the id of the token;
    \item $Node\_ID$: the id of the hosting node;
    \item $Node\_Value$: the value of the hosting node;
    \item $Arc\_Time\_Delay$: the time delay for the causal-effect event between the hosting node and the target node to be triggered; and
    \item $Arc\_Weight$: the arc weight between the hosting node and the target node, for possible weighted node value updates. 
\end{enumerate}

The $Token\_ID$ is a unique number identifying the token, which remains unchanged throughout the iteration process. The $Node\_ID$ and $Node\_Value$ are the id and value of the node where the token is located. When the token enters a new arc, the two attribute values of the token, $Arc\_Time\_Delay$ and $Arc\_Weight$, will be updated to the time delay and weight of the new arc. When the token enters a new node, the node will be activated, and the node value will be updated using the following formula:
\begin{equation}\label{eq:node-value-update}
    \widetilde{A_i}=f(A_i+\sum_{j=1}^{n}Token_{Arc\_Weight}^jToken_{Node\_Value}^j)
\end{equation}
where $n$ represents the number of tokens that arrive at node $i$. $Token_{Arc\_Weight}^j$ represents the $Arc\_Weight$ of the arc passed by $Token^j$, and $Token_{Node\_Value}^j$ represents the $Node\_Value$ of the starting node of $Token^j$. $A_i$ and $\widetilde{A_i}$ represent the value of the node before and after the update in the current iteration. After the update is completed, $Node\_ID$ and $Node\_Value$ of the token will also be updated to take the id and the new value of the node arriving.

\begin{figure}[t]
    \centering
    \subfigure[]{%
    \resizebox*{3cm}{!}{\includegraphics{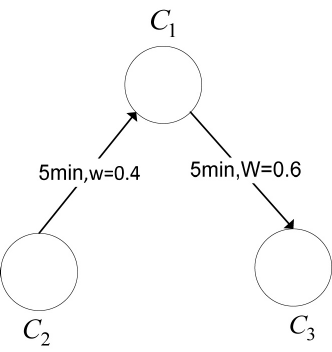}}}\hspace{5pt}
    \subfigure[]{%
    \resizebox*{3cm}{!}{\includegraphics{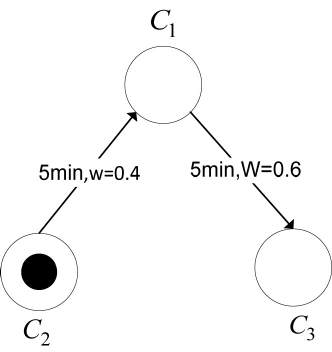}}}\hspace{5pt}
    \subfigure[]{%
    \resizebox*{3cm}{!}{\includegraphics{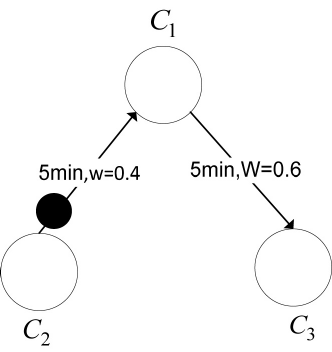}}}\hspace{5pt}
    \subfigure[]{%
    \resizebox*{3cm}{!}{\includegraphics{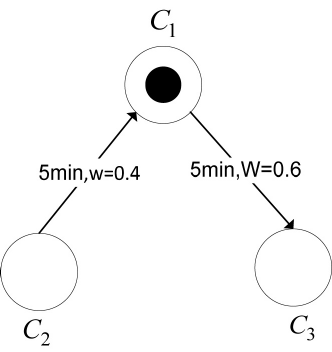}}}\hspace{5pt}
    \subfigure[]{%
    \resizebox*{3cm}{!}{\includegraphics{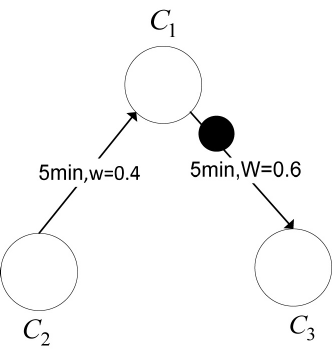}}}\hspace{5pt}
    \subfigure[]{%
    \resizebox*{3cm}{!}{\includegraphics{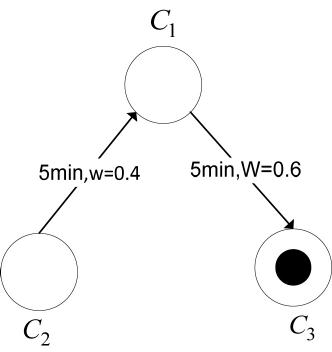}}}\hspace{5pt}
    \caption{The example of token behavior.}
    \label{fig:example-of-token-behavior}
\end{figure}

To better understand the process of a token's attribute change, we provide an example as follows. Given the Token-FCM shown in Figure~\ref{fig:example-of-token-behavior}a, the three nodes' initial values: $\{C_1:0.5, C_2:0.6, C_3:0.7\}$, and the threshold function $Sigmoid()$, the update process includes the following:
\begin{itemize}
  \item [1)] 
  When the token is at the $C_2$ node, its attributes are:$\{Token\_ID:1, Node\_ID:2, Node\_Value:0.6, Arc\_Time\_Delay:\sim, Arc\_Weight:\sim\}$, as shown in Figure~\ref{fig:example-of-token-behavior}b.     
  \item [2)]
  When the iteration starts, the token enters the arc: $C_2 \rightarrow C_1$, then the token's attributes change to:$\{Token\_ID:1, Node\_ID:2, Node\_Value:0.6, Arc\_Time\_Delay:5, Arc\_Weight:0.4\}$, as shown in Figure~\ref{fig:example-of-token-behavior}c.
  \item [3)]
  After 5 minutes, the token arrives at the $C_1$ node. Then immediately $C_1$ is activated and the node value is updated using Eq.~\ref{eq:node-value-update}. The token's attributes become: $\{Token\_ID:1, Node\_ID:1, Node\_Value:0.677, Arc\_Time\_Delay:\sim, Arc\_Weight:\sim \}$, as shown in Figure~\ref{fig:example-of-token-behavior}d.
  \item [4)]
  As the iteration continues, the token enters the arc $C_1 \rightarrow C_3$ and changes its attributes to: $\{Token\_ID:1, Node\_ID:1, Node\_Value:0.677, Arc\_Time\_Delay:5, Arc\_Weight:0.6\}$, as shown in Figure~\ref{fig:example-of-token-behavior}e.
  \item [5)]
  After 5 minutes, the token arrives at the node $C_3$ and subsequently activates it. Then the token's attributes are changed to: $\{Token\_ID:1, Node\_ID:3, Node\_Value:0.7514, Arc\_Time\_Delay:\sim, Arc\_Weight:\sim\}$, as shown in Figure~\ref{fig:example-of-token-behavior}f. 
\end{itemize}

\subsection{Token-FCM iterations}
\label{sec:fcm-iteration}
The pseudocode of the Token-FCM method is given in \textbf{Algorithm}~\ref{al-fcm}. The input includes the nodes' initial values $\{Node_i\}_{i=1}^n$, FCM arc weights $\{w_{ij}\}_{i=1,j=1}^n$, the corresponding time delay $\{time\_delay_{ij}\}_{i=1,j=1}^n$, the maximum iteration time $T$, and the step size $t$. The result of the process is a vector of nodes $\{Node_i^*\}_{i=1}^n$ that represent the ultimate value of the FCM. The algorithm's main procedures are:
\begin{enumerate}
  \item FCM initialization (Line 1 of \textbf{Algorithm}~\ref{al-fcm}): Initialize the value of each node and generate the corresponding token.    
\item Token movement (Lines 3-6 of \textbf{Algorithm}~\ref{al-fcm}): For all nodes that contain tokens, move their tokens to the arc according to the behavior defined in Section ~\ref{sec:token-behavior-design}.
  \item  Processing of time delays (Lines 7-10 of \textbf{Algorithm}~\ref{al-fcm}): Time delays are performed on all tokens and each iteration reduces the remaining time delay by $t$. For those having zero remaining time, move them to the target node.
  \item Update of the node value (Lines 11-13 of \textbf{Algorithm}~\ref{al-fcm}): For all activated nodes, update their node values according to Eq.~\ref{eq:node-value-update}. 
  \item Return to step (2) unless $T\leq0$. 
\end{enumerate}

\begin{algorithm}[t]
\caption{The simulation of Token-FCM}
\hspace*{0.02in} {\bf Input:} 
$Node_i$, $w_{ij}$, $time\_delay_{ij}$, $T$, $t$\\
\hspace*{0.02in} {\bf Output:} 
Stabilized FCM node vector $Node_i^*,(i=1,...,n)$ 
\begin{algorithmic}[1]
\State $Node_i.isActivate=true,(i=1,…,n)$ (Activate all nodes)
\While{$T>0$}
    \For{each node \textbf{in} FCM}
        \If{$Node_i.isActivate==true$}
            \State Move the token according to the Sec3.2.
        \EndIf
    \State $Node_i.isActivate=false$ 
    \EndFor
    \For{each token \textbf{in} FCM arc}
        \State $Token.time\_delay=Token.time\_delay-t$
        \If{$Token.time\_delay==0$}
            \State Enter the target node and activate it.
        \EndIf
    \EndFor
    \For{each node \textbf{in} FCM}
        \If{$Node_i.isActivate==true$}
            \State Calculate the updated node value by Eq.2.
        \EndIf
    \EndFor
    \State $T=T-t$ 
\EndWhile
\State \textbf{Return} $Node_i^*$,(i=1,…,n)
\end{algorithmic}
\label{al-fcm}
\end{algorithm}

\section{Token-FCM initialization}
\label{sec:FCM-initialization}
As already noted, Token-FCM's effectiveness is highly dependent on the initialization of nodes, i.e., the initial risk value of each node's corresponding component of the design under study. To this end, this section presents a new initialization method built mainly upon linguistic term sets and group decision-making. This method consists of two steps: (1) Collecting expert opinions based on the group decision-making method and the linguistic term sets; and (2) Deriving the initial value of the FCM node using the RPN calculation method based on the linguistic term sets. The details are given in the following subsections.

\subsection{Collecting expert opinions}
In engineering design, each design component has a risk of failure due to mechanical failures, environmental impacts, and other reasons. To accurately quantify the risk index of individual components (i.e., the initial values of Token-FCM nodes), we first make use of the risk priority number (RPN) calculation method from the FMEA domain. RPN is evaluated through the comprehension of fault occurrence (O), fault severity (S), and fault detection difficulty (D)~\cite{wang2009risk}.

Firstly, multiple experts give the occurrence of the fault (O), the severity of the fault (S), and the difficulty of the fault detection (D) of the design based on their own experience and knowledge. Tables 1-3 display the linguistic term sets and meanings of different O, S, and D risk indices, which are adapted from~\cite{wang2009risk}.

However, for complex engineering products, personal experience and knowledge are often very limited, which can easily lead to insufficient understanding of the design and result in biased decision-making~\cite{wang2021decision}. Therefore, group Decision Making (GDM) is used in this paper to improve the RPN calculation. The detailed procedures are as follows. After gathering expert opinions, we express them in the form of a collective opinion using the following formula: 

\begin{subequations} \label{subeqnexample}
\begin{equation}
     C_O=\left\{\frac{k_{very\ low}^O}{K_O},\frac{k_{low}^O}{K_O},\frac{k_{medium}^O}{K_O},\frac{k_{high}^O}{K_O},\frac{k_{very\ high}^O}{K_O}\right\}
\end{equation}
\begin{equation}
     C_S=\left\{\frac{k_{very\ low}^S}{K_S},\frac{k_{low}^S}{K_S},\frac{k_{medium}^S}{K_S},\frac{k_{high}^S}{K_S},\frac{k_{very\ high}^S}{K_S}\right\}
\end{equation}
\begin{equation}
    C_D=\left\{\frac{k_{very\ low}^D}{K_D},\frac{k_{low}^D}{K_D},\frac{k_{medium}^D}{K_D},\frac{k_{high}^D}{K_D},\frac{k_{very\ high}^D}{K_D}\right\}
\end{equation}
\label{eq:expert-opinion}
\end{subequations}
where $K_O$, $K_S$ and $K_D$ are the number of experts participating in the evaluation, $k_j^i$($i=O,S,D$, $j$=\textit{very low, very low, medium, high, very high}) is the number of experts who choose the $j$-th risk rating for the $i$-th risk index. 

In order to facilitate the calculation of $C_O$, $C_S$, $C_D$, Eq.~\ref{eq:expert-opinion} is rewritten in the form of probabilistic linguistic term sets (PLTs)~\cite{pang2016probabilistic}:
\begin{subequations} \label{subeqnexample}
\begin{equation}
    C_O=\left\{s_{-2}(p_{-2}^O),s_{-1}(p_{-1}^O),s_0(p_0^O),s_1(p_1^O),s_2(p_2^O)\right\},
\end{equation}
\begin{equation}
     C_S=\left\{s_{-2}(p_{-2}^S),s_{-1}(p_{-1}^S),s_0(p_0^S),s_1(p_1^S),s_2(p_2^S)\right\},
\end{equation}
\begin{equation}
   C_D=\left\{s_{-2}(p_{-2}^D),s_{-1}(p_{-1}^D),s_0(p_0^D),s_1(p_1^D),s_2(p_2^D)\right\}
\end{equation}
\label{eq:expert-opinion-plt}
\end{subequations}
where $S=\{s_{-2}=very low,\ s_{-1}=low,\ s_0=medium,$ 
$s_1=high,s_2=very\ high\}, p_j^i=\frac{k_j^i}{K_j} \ (i=O,S,D,j=-2,\ldots,0,\ldots,2)$.

\begin{table}
\caption{Failure probability and corresponding linguistic variables.}
\centering
\begin{tabular}{lcccccc} \hline
Rating      & Failure Probability              \\ \hline
Very   high & 1:20\textless{}=failure rate\textless{}1:2     \\ 
High        & 1:200\textless{}=failure rate\textless{}1:20    \\
Medium      & 1:10000\textless{}=failure rate\textless{}1:200  \\
Low         & 1:20000\textless{}=failure rate\textless{}1:10000 \\
Very   low  & failure rate\textless{}1:20000                    \\ \hline
\end{tabular}
\label{table1}
\end{table}

\begin{table}
\caption{Detection and corresponding linguistic variables.}
\centering
\begin{tabular}{lcccccc} \hline
Rating      & Difficulty of detection  \\ \hline
Very   high & Hard to detect   \\ 
High        & Very low chance to detect \\
Medium      & Moderate chance to detect \\
Low         & High chance to detect \\
Very   low  & Very High chance to detect  \\ \hline
\end{tabular}
\label{table3}
\end{table}

\begin{table}[t]
\caption{Severity and corresponding linguistic variables.}
\centering
\resizebox{\linewidth}{!}{
\begin{tabular}{lcccccc} \hline
Rating      & Failure effect  \\ \hline
Very   high & Cause human death or system damage   \\ 
High        & Cause great injuries, economic loss or failed functions  \\
Medium      & Cause minor injuries, economic loss or task delay  \\
Low         & No injuries, just some economic loss or task delay \\
Very   low  & Almost no injuries, economic loss, and task delay   \\ \hline
\end{tabular}
}
\label{table2}
\end{table}

\subsection{Deriving initial values of the FCM}

After obtaining a collective and quantified opinion about the design's risks, we use the RPN calculation method to obtain the final initial value for running Token-FCM. The original RPN calculation method is as follows~\cite{wang2009risk}: 
\begin{equation}
   RPN=O\times S\times D 
   \label{eq:ori-rpn-cal}
\end{equation}

To make it applicable to the PLTs in Eq.~\ref{subeqnexample}, we rewrite it to the following form:

\begin{flalign}
\begin{array}{ll}
RPN=C_O^{w_o}\otimes{C_S}^{w_s}\otimes{C_D}^{w_d}=\\
g^{-1}\left( \bigcup\left(g\left(s_k\right)^{W_O}g\left(s_i\right)^{w_S}g\left(s_j\right)^{w_D}\right)\left(p_k^Op_i^Sp_j^D\right) \right) 
\end{array}&
\label{eq:plt-rpn-cal}
\end{flalign}

where $w_o, w_s, w_d$ represent the weights assigned to various risk indices, totaling 1 (i.e., $w_o+w_s+w_d=1$). These weights are determined through expert judgment and empirical evidence. They are adaptable and can be modified to suit particular system circumstances. $g()$ and $g^{-1}()$ are the conversion function and the inverse function of linguistic variables and constants~\cite{montserrat2018consensus}, respectively:

\begin{subequations} \label{subeqnexample}
\begin{equation}
    g\left(s_i\right)=\frac{i}{2t}+\frac{1}{2},
\end{equation}
\begin{equation}
    g^{-1}\left(i\right)=s_{(2i-1)t},
\end{equation}
\label{eq:g-cal}
\end{subequations}

where $t$ is the range of linguistic terms in the $S=\{s_i|i=-t,...,0,...,t\}$.

Obviously, the RPN calculated by Eq.~\ref{eq:plt-rpn-cal} is still a set of fuzzy terms in the form of PLTs. To facilitate subsequent FCM iterations, the RPN must be defuzzified, that is, converting the RPN in the form of PLT: $\left\{s_{k_1}\left(p_{k_1}\right),\ldots,s_{k_i}\left(p_{k_i}\right),\ldots,s_{k_n}\left(p_{k_n}\right)\right\}$  to the following values:
\begin{equation}
    RPN=\sum_{i=1}^{n}{k_i\times p_{k_i}}
\label{eq:plt-dufuzz}
\end{equation}

Now we have obtained the initial value for each node in the Token-FCM, in terms of a collective and quantitative expert opinion on design risks.

\section{The overall risk assessment method}
\label{sec:overallmethod}
Based on the preceding Token-FCM method and its initialization method, we propose a systematic decision support method for risk assessment and key risk prediction of engineering designs with two-way, dynamic causal-effect relations between risks. It consists of three main steps:
\begin{enumerate}
    \item [1)] \textbf{Token-FCM initialization}: calculate Token-FCM nodes' initial values using Eqs.~\ref{eq:plt-rpn-cal} and \ref{eq:plt-dufuzz}.
  \item [2)] \textbf{Token-FCM iteration and design risk evaluation}: run \textbf{Algorithm~\ref{al-fcm}} to obtain each node's final value, i.e., the assessed risk values of the design's components.
  \item [3)] \textbf{Decision making on design risks}: obtain a comprehensive assessment and decision-making on the design's risks based on evaluated risk values in Step 2.
\end{enumerate}

\subsection{Token-FCM initialization}
Firstly, we gather the opinions of experts at each FCM node and use the method described in Section ~\ref{sec:FCM-initialization} to compute the results, then the initial Token-FCM node value vector is obtained, denoted as: 
\begin{equation*}
C^0=\left\{RPN_1,\cdots,RPN_i,\cdots,RPN_n\right\}
\label{eq:node-vec}
\end{equation*}

\subsection{Token-FCM iteration and design risk evaluation}
The causal-effect relations between design risks in the system are analyzed, and then experts can establish the FCM of the risks in the system and assign the corresponding weights and time delay to different arcs. In this step, experts should reach a consensus on the direction of the causal-effect relation, the weight of the arc, and the time delay of the causal-effect event.

The minimum iteration time unit $t$ and the total iteration time $T$ are determined according to the time delay of each arc in the Token-FCM. Then the Token-FCM is processed according to \textbf{Algorithm \ref{al-fcm}} in Section ~\ref{sec:fcm-iteration} to obtain the final node value vector, that is, the Dynamic RPN: $DRPN_i$ for each design risk considering the risk of the run time of the overall system.

\begin{equation*}
    C^\ast=\left\{DRPN_1,\ldots,DRPN_i,\ldots,DRPN_n\right\}
\label{eq:node-vec-final}
\end{equation*}

In order to analyze the interaction relation between design risks more comprehensively, it is necessary to separately evaluate the impact of each risk on other risks.
Each node is activated separately, and the values of other nodes are set to 0, forming n initial node vectors:

\begin{equation*}
    C_1=\left\{RPN_1,…,0,…,0\right\}
\end{equation*}
\begin{equation*}
    …
\end{equation*}    
\begin{equation*}
    C_i=\left\{0,…,RPN_i,…,0\right\}
\end{equation*}
\begin{equation*}
    …
\end{equation*}
\begin{equation*}
    C_n=\left\{0,…,0,…,RPN_n\right\}
\end{equation*}

The above n initial vectors are used as the initial state of FCM for n iterations, and the n FCM simulation vectors are obtained:

\begin{equation*}
    C_1^*=\left\{DRPN_{11},…,DRPN_{1i},…,DRPN_{1n}\right\}
\end{equation*}
\begin{equation*}
    …
\end{equation*}    
\begin{equation*}
    C_i^*=\left\{DRPN_{i1},…,DRPN_{ii},…,DRPN_{in}\right\}
\end{equation*}
\begin{equation*}
    …
\end{equation*}
\begin{equation*}
    C_n^*=\left\{DRPN_{n1},…,DRPN_{ni},…,DRPN_{nn}\right\}
\end{equation*}

These vectors signify the effect of one design risk on other risks. They are also an essential index for risk assessment. The magnitude of the influence of design risk $DR_1$ on the other n-1 design risks is represented by vector $C_1^*$, with the design risk having the highest value in vector $C_1^*$ being the one most affected by design risk $DR_1$.

\subsection{Decision making on design risk}
Through the above two steps, we calculate an initial design risk index (the risk of the system itself), a final design risk index (the risk that is taken into account when the system is in operation), and the degree to which the risk influences affects other risks. Based on these three indicators, a systematic analysis table is formed, as shown in Table ~\ref{table:System-risk-analysis}, which is a example product design risk analysis table.

\begin{table}[htb]
\caption{Design risk analysis table.}
\centering
\resizebox{\linewidth}{!}{
\begin{tabular}{lcccccc} \hline
No.     & Design risk Name & RPN & DRPN & Most Impact DR\\ \hline
$DR_1$ & Valve jam & 0.56  & 0.77  & $DR_i$ \\ 
\dots   &  \dots  & \dots &\dots &\dots     \\
$DR_i$  &  Battery failure & 0.45  & 0.68  & $DR_n$ \\
\dots      &  \dots  & \dots &\dots &\dots   \\
$DR_n$  & Pipeline blockage  & 0.87  & 0.92  & $DR_3$   \\ \hline
\end{tabular}
}
\label{table:System-risk-analysis}
\end{table}

Based on this table, decision makers and experts can comprehensively investigate design risks, quantify different risk indexes, and predict key systems, making it easier to formulate design plans and improve product reliability.

\section{Tha case study}
\label{sec:results}
In this section, we apply the proposed Token-FCM method to a diesel engine system as a case study to demonstrate its applicability and potential for assessing system risks.

Initially, this approach evaluates the static risk of the system using the risk priority number (RPN). This number represents the severity, occurrence rate, and detection rate of potential failures of each system component, thus quantifying the risk index for individual component failures. Additionally, it assesses dynamic risk through the dynamic risk priority number (DRPN). Building upon the RPN, the DRPN takes into account the intricate interactions and mutual influences of component failures during system operation, allowing for a thorough evaluation of the dynamic risk index for each component during system usage. By employing these two indices, designers gain the ability to conduct a comprehensive risk analysis and evaluation of the system, facilitating proactive design optimization measures.

\subsection{Token-FCM initialization}

\begin{figure}[htb]
    \centering
    \resizebox*{8cm}{!}{\includegraphics{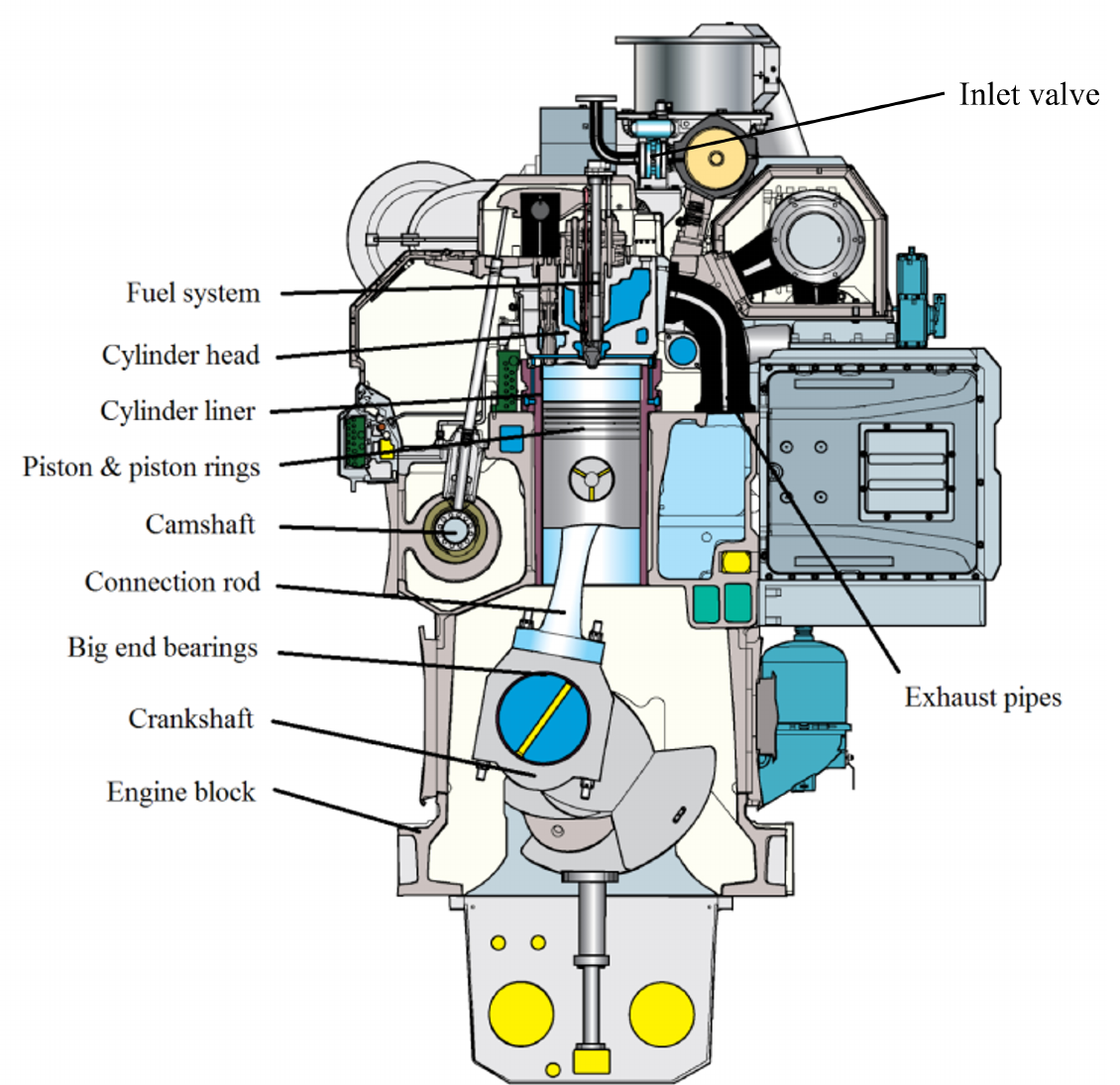}}
    \caption{The cross-section of diesel engine \cite{docent2016approximating}.}
    \label{fig:diesel-engine}
\end{figure}

An annotated cross section of the diesel engine is shown in Figure ~\ref{fig:diesel-engine}. The diesel engine occasionally has some serious design risks during operation, mainly including: ``$DR_1$: inlet valve failure'', ``$DR_2$: piston failure'', ``$DR_3$: cylinder head cracking'', ``$DR_4$: fuel injector jam'', ``$DR_5$: big end bearing failure'', and ``$DR_6$: camshaft failure'', and these design risks will affect each other unidirectionally and bidirectionally.

According to Section \ref{sec:overallmethod}, first initialize the value for each FCM node, that is, for each design risk index. To obtain a more reliable risk index, subjective evaluations of 20 experts are collected on the O, S, and D indexes of each design risk. All evaluations are expressed by linguistic term sets: $S=\{s_{-2}=very\ low,\ \ s_{-1}=low,\ \ s_0=medium,\ \ s_1=high,$ 
$s_2=very\ high\}$, as shown in Table ~\ref{table:Opinion-of-RPN}. The weight of each risk index is determined through an expert evaluation, with $W_O$ being 0.5, $W_S$ being 0.35, and $W_D$ being 0.15.

\begin{table}
\caption{The opinion of experts in risk index.}
\centering
\resizebox{\linewidth}{!}{
\begin{tabular}{cccccc|ccccc|ccccc} \hline
\multirow{2}{*}{DR no.} & \multicolumn{5}{c}{O}  & \multicolumn{5}{c}{S}  & \multicolumn{5}{c}{D}  \\ \cline{2-16} 
                        & $s_{-2}$ & $s_{-1}$ & $s_0$ & $s_1$ & $s_2$ & $s_{-2}$ & $s_{-1}$ & $s_0$ & $s_1$ & $s_2$ & $s_{-2}$ & $s_{-1}$ & $s_0$ & $s_1$ & $s_2$ \\  \hline
$DR_1$                     & 3  & 5  & 10 & 1  & 1  & 0  & 2  & 9  & 6  & 3  & 0  & 2  & 6  & 7  & 5  \\ 
$DR_2$                     & 1  & 2  & 3  & 9  & 5  & 0  & 1  & 5  & 9  & 5  & 3  & 8  & 7  & 2  & 0  \\
$DR_3$                      & 1  & 3  & 10 & 5  & 1  & 12 & 7  & 1  & 0  & 0  & 4  & 6  & 8  & 1  & 1  \\
$DR_4$                      & 0  & 4  & 9  & 7  & 0  & 13 & 5  & 1  & 1  & 0  & 10 & 6  & 3  & 1  & 0  \\
$DR_5$                    & 9  & 8  & 2  & 1  & 0  & 0  & 10 & 9  & 1  & 0  & 2  & 4  & 8  & 5  & 1  \\
$DR_6$                     & 0  & 5  & 8  & 4  & 3  & 0  & 5  & 8  & 5  & 2  & 13 & 4  & 3  & 0  & 0 \\ \hline
\end{tabular}
}
\label{table:Opinion-of-RPN}
\end{table}

The opinions of experts are expressed by PLTs:
\begin{tiny}
\begin{equation*}
DR_1=
    \begin{Bmatrix}
 & C_O=\left\{s_{-2}(0.15),s_{-1}(0.25),s_0(0.5),s_1(0.05),s_2(0.05)\right\},\\ 
 & C_S=\left\{s_{-1}(0.1),s_0(0.45),s_1(0.3),s_2(0.15)\right\}, \\ 
 & C_D=\left\{s_{-1}(0.1),s_0(0.3),s_1(0.35),s_2(0.25)\right\}
\end{Bmatrix}
\end{equation*}
\begin{equation*}
DR_2=
    \begin{Bmatrix}
 & C_O=\left\{s_{-2}(0.05),s_{-1}(0.1),s_0(0.15),s_1(0.45),s_2(0.25)\right\},\\ 
 & C_S=\left\{s_{-1}(0.05),s_0(0.25),s_1(0.45),s_2(0.25)\right\}, \\ 
 & C_D=\left\{s_{-2}(0.15),s_{-1}(0.4),s_0(0.35),s_1(0.1)\right\}
\end{Bmatrix}
\end{equation*}
\begin{equation*}
DR_3=
    \begin{Bmatrix}
 & C_O=\left\{s_{-2}(0.05),s_{-1}(0.15),s_0(0.5),s_1(0.25),s_2(0.05)\right\},\\ 
 & C_S=\left\{s_{-2}(0.6),s_{-1}(0.35),s_0(0.05)\right\}, \\ 
 & C_D=\left\{s_{-2}(0.2),s_{-1}(0.3),s_0(0.4),s_1(0.05),s_2(0.05)\right\}
\end{Bmatrix}
\end{equation*}\begin{equation*}
DR_4=
    \begin{Bmatrix}
 & C_O=\left\{s_{-1}(0.2),s_0(0.45),s_1(0.35)\right\},\\ 
 & C_S=\left\{s_{-2}(0.65),s_{-1}(0.25),s_0(0.05),s_1(0.05)\right\}, \\ 
 & C_D=\left\{s_{-2}(0.5),s_{-1}(0.3),s_0(0.15),s_1(0.05)\right\}
\end{Bmatrix}
\end{equation*}\begin{equation*}
DR_5=
    \begin{Bmatrix}
 & C_O=\left\{s_{-2}(0.45),s_{-1}(0.4),s_0(0.1),s_1(0.05)\right\},\\ 
 & C_S=\left\{s_{-1}(0.5),s_0(0.45),s_1(0.05)\right\}, \\ 
 & C_D=\left\{s_{-2}(0.1),s_{-1}(0.2),s_0(0.4)s_1(0.25),s_2(0.05)\right\}
\end{Bmatrix}
\end{equation*}\begin{equation*}
DR_6=
    \begin{Bmatrix}
 & C_O=\left\{s_{-1}(0.25),s_0(0.4),s_1(0.2),s_2(0.15)\right\},\\ 
 & C_S=\left\{s_{-1}(0.25),s_0(0.4),s_1(0.25),s_2(0.1)\right\}, \\ 
 & C_D=\left\{s_{-2}(0.65),s_{-1}(0.2),s_0(0.15)\right\}
\end{Bmatrix}
\end{equation*}
\end{tiny}

Calculate the static risk index $RPN_i$ of each design risk using Eq.~\ref{eq:plt-rpn-cal}, and perform normalization processing~\cite{pang2016probabilistic}, that is, the sum of probabilities $p$ in the PLTs is normalized to 1.

\begin{tiny}
\begin{equation*}
RPN_1=\left\{s_{-2}(0.08),s_{-0.59}(0.12),s_0(0.04),s_{0.13}(0.74),s_{1.14}(0.02)\right\}
\end{equation*}
\begin{equation*}
RPN_2=\left\{s_{-2}(0.09),s_{-0.73}(0.03),s_0(0.01),s_{0.21}(0.8),s_{1.26}(0.06)\right\}
\end{equation*}
\begin{equation*}
RPN_3=\left\{s_{-2}(0.62),s_{-0.59}(0.18),s_0(0.02),s_{0.04}(0.18)\right\}
\end{equation*}
\begin{equation*}
RPN_4=\left\{s_{-2}(0.76),s_{-0.59}(0.11),s_{0.08}(0.13)\right\}
\end{equation*}
\begin{equation*}
RPN_5=\left\{s_{-2}(0.53),s_{-0.89}(0.33),s_0(0.02),s_{0.13}(0.12)\right\}
\end{equation*}
\begin{equation*}
RPN_6=\left\{s_{-2}(0.43),s_{-1}(0.06),s_0(0.02),s_{0.08}(0.49),s_{1.12}(0.01)\right\}
\end{equation*}
\end{tiny}

Then, calculate the final static risk index $RPN_i$ for each design risk defuzzification by Eq.~\ref{eq:plt-dufuzz}, which is the initial node value of FCM:
\begin{small}
\begin{equation*}
RPN_1:-0.1118,\ RPN_2:0.0417,\ RPN_3:-1.3390\ 
\end{equation*}
\begin{equation*}
RPN_4:-1.5745,\ RPN_5:-1.3381,\ RPN_6:-0.8696
\end{equation*}
\end{small}

\subsection{Token-FCM iteration and design risk evaluation}
\begin{figure}[htb]
    \centering
    \resizebox*{7cm}{!}{\includegraphics{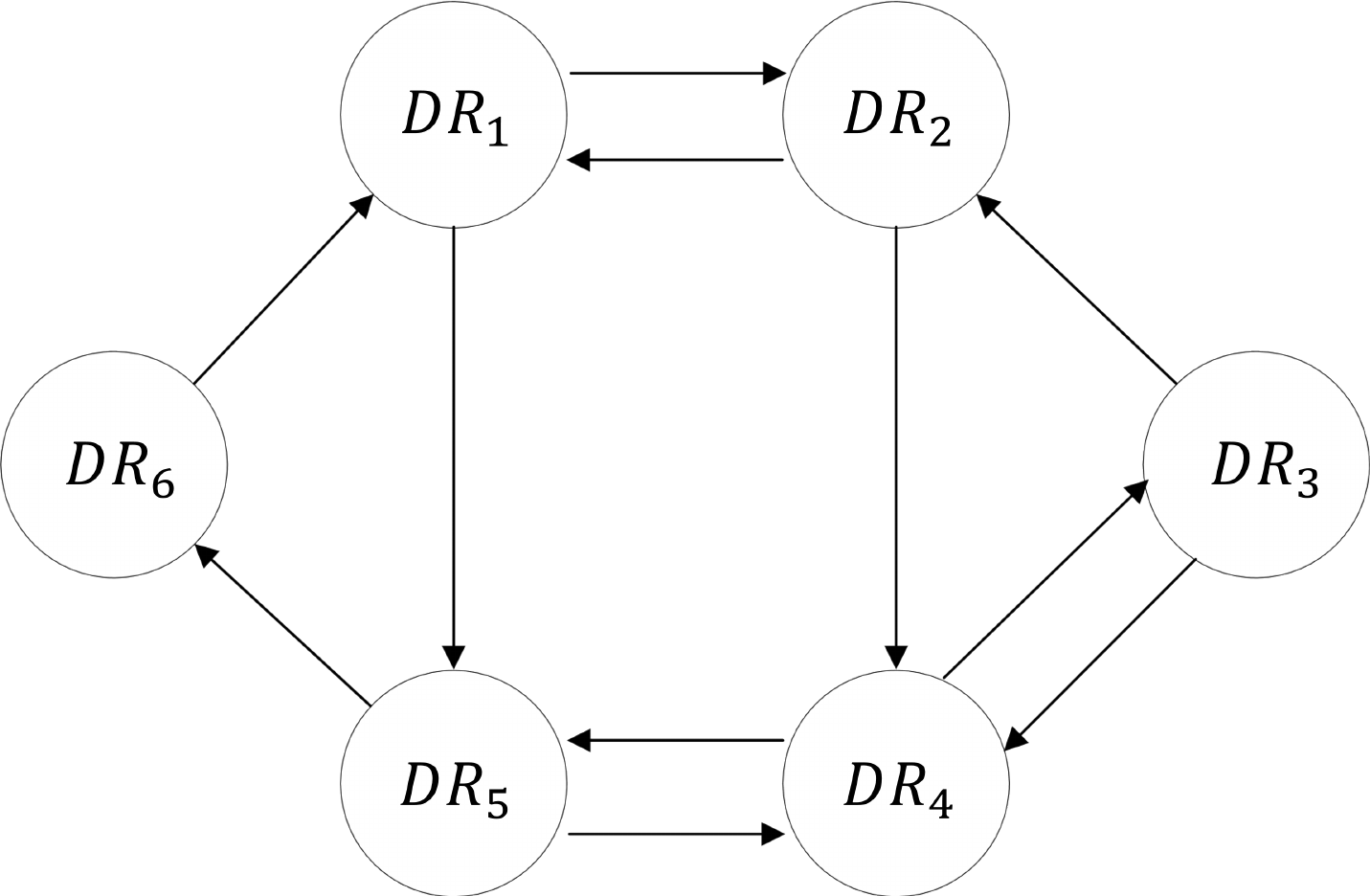}}
    \caption{FCM of the diesel engine design risk.}
    \label{fig:FCM-system-risk}
\end{figure}

Through the design risk of the diesel engine and the analysis of causal-effect events between them, for example, inlet valve failure ($DR_1$) will cause piston failure ($DR_2$), fuel injector jam ($DR_4$) will cause inlet valve failure ($DR_1$), piston failure ($DR_2$) will cause the injector ($DR_4$) to be blocked, etc., construct the FCM as shown in Figure \ref{fig:FCM-system-risk}.

After expert review and empirical data analysis, the impact weight of each design risk and the time delay triggered by each design risk event are obtained, as shown in Table ~\ref{table:impact-relation}.

\begin{table}[htb]
\caption{The impact relation and time delay between design risks.}
\centering
\begin{tabular}{cccc} \hline
\multicolumn{2}{c}{design risk and their influence} & $w_{ij}$       & Time delay(min) \\ \hline
\multirow{2}{*}{$DR_1\rightarrow $}               & $DR_2$              & 0.8 & 2        \\
                                   & $DR_5$              & 0.2       & 10       \\
\multirow{2}{*}{$DR_2\rightarrow $}               & $DR_1$              & 0.6      & 2        \\
                                   & $DR_4$              & 0.8 & 4        \\
\multirow{2}{*}{$DR_3\rightarrow $}               & $DR_2$              & 0.4    & 4        \\
                                   & $DR_4$              & 0.8 & 2        \\
\multirow{2}{*}{$DR_4\rightarrow $}               & $DR_3$              & 0.4    & 4        \\
                                   & $DR_5$              & 0.6 & 4        \\     
\multirow{2}{*}{$DR_5\rightarrow $}               & $DR_4$              & 0.6    & 4        \\
                                   & $DR_6$              & 0.6      & 4        \\
\multirow{2}{*}{$DR_6\rightarrow $}               & $DR_1$              & 0.4    & 4        \\
                                   & $DR_2$              & 0.8 & 4      \\ \hline
\end{tabular}
\label{table:impact-relation}
\end{table}

\begin{figure}[htb]
    \centering
    \resizebox*{8cm}{!}{\includegraphics{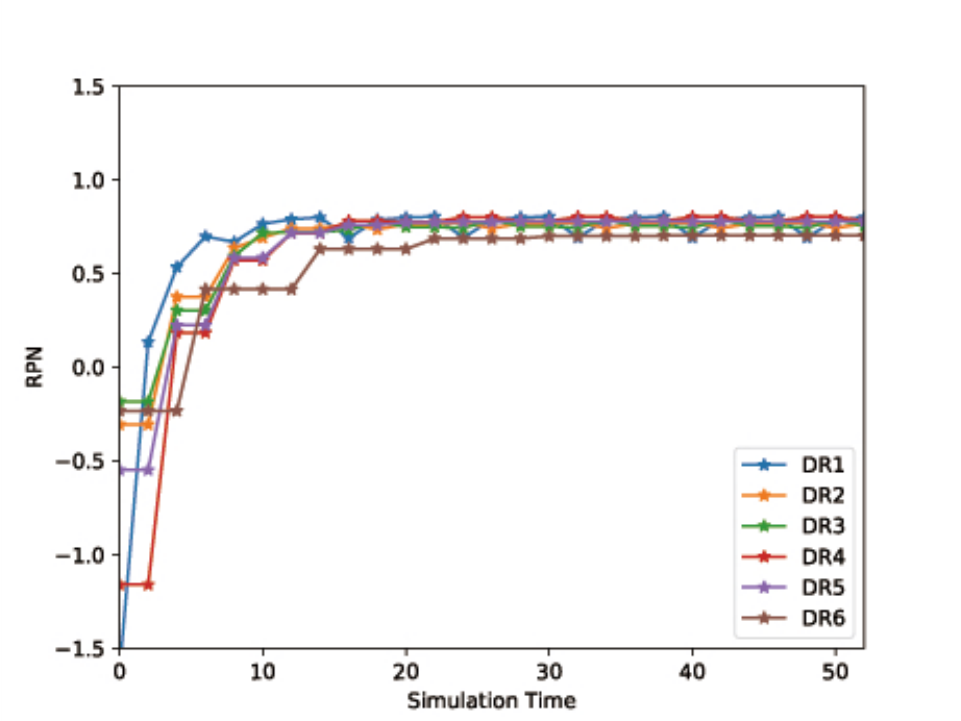}}
    \caption{Changes in risk index $RPN_i$ of design risks during iteration.}
    \label{fig:Changes-in-RPN-index}
\end{figure}

Next, according to Section \ref{sec:overallmethod}, the FCM starts the iteration process, the time interval of each iteration $t$ is 2 minutes, and the overall iteration time $T$ is 50 minutes. In the iteration process, the change in the risk index $RPN_i$ for each design risk is shown in Figure ~\ref{fig:Changes-in-RPN-index}.

When the simulation time reaches 40 minutes, it is observed that the $RPN_i$ index of all design risks has generally stabilized: Some index values no longer change, and other index values enter a finite cycle. Obviously, it meets the definition of a steady-state FCM~\cite{papageorgiou2013fuzzy}.

The specific changes in the $RPN_i$ index of each design risk with the simulation time are shown in the Table \ref{table:specific-value-change} below.

\begin{table*}[htb]
\caption{The specific value of $RPN_i$ index of design risks during iteration.}
\centering
\begin{tabular}{ccccccc} \hline
Time & $DR_1$   & $DR_2$  & $DR_3$   & $DR_4$   & $DR_5$   & $DR_6$   \\ \hline
0    & -0.1118 & 0.0417 & -1.3390 & -1.5745 & -1.3381 & -0.8696 \\
2    & 0.4623  & 0.3401 & -1.3390 & 0.0648 & -1.3381 & -0.8696 \\
4    & 0.5805  & 0.5067 & 0.0878  & 0.5262  & -1.3381 & 0.1635  \\
6   & 0.6971   & 0.3753 & 0.3029  & 0.1840   & 0.2236  & 0.4151       \\
8   & 0.6676  & 0.6358 & 0.5930  & 0.5699  & 0.5827  & 0.4151  \\
10  & 0.7643 &  0.6904 &  0.7161 &  0.5699 &  0.5827 &  0.4151  \\
12  &  0.7886 &  0.74 &  0.7199 &  0.715 &  0.716 &  0.4151 \\
14  &  0.7991 &  0.74 &  0.7199 &  0.715 &  0.716 &  0.6299  \\
16  &  0.6898 &  0.7635 &  0.7322 &  0.7796 &  0.7586 &  0.6299  \\
18   & 0.8177  & 0.9009 & 0.7719  & 0.8085  & 0.2087  & 0.5717  \\
20   & 0.7955  & 0.8256 & 0.78    & 0.8115  & 0.5901  & 0.5717  \\
22  &  0.8036 &  0.7638 &  0.7478 &  0.7746 &  0.7732 &  0.686 \\
24  &  0.6907 &  0.7707 &  0.7422 &  0.7998 &  0.7752 &  0.686 \\
26  &  0.7871 &  0.7398 &  0.7843 &  0.7998 &  0.7752 &  0.686 \\
28  &  0.7988 &  0.7659 &  0.7511 &  0.7799 &  0.7782 &  0.686 \\
30  &  0.804 &  0.7659 &  0.7511 &  0.7799 &  0.7782 &  0.6987\\
32  &  0.6908 &  0.7714 &  0.7433 &  0.8014 &  0.7766 &  0.6987 \\
34  &  0.7872 &  0.741 &  0.7862 &  0.8014 &  0.7766 &  0.6987 \\
36  &  0.799 &  0.7662 &  0.7515 &  0.7803 &  0.7786 &  0.6987 \\
38  &  0.8041 &  0.7662 &  0.7515 &  0.7803 &  0.7786 &  0.7014\\
40  &  0.6908 &  0.7716 &  0.7434 &  0.8016 &  0.7767 &  0.7014 \\
42  &  0.7872 &  0.7412 &  0.7866 &  0.8016 &  0.7767 &  0.7014 \\
44  &  0.799 &  0.7662 &  0.7516 &  0.7803 &  0.7786 &  0.7014 \\
46  &  0.8041 &  0.7662 &  0.7516 &  0.7803 &  0.7786 &  0.702 \\
48  &  0.6908 &  0.7716 &  0.7434 &  0.8016 &  0.7767 &  0.702 \\
50  &  0.7872 &  0.7412 &  0.7867 &  0.8016 &  0.7767 &  0.702  \\ \hline
\end{tabular}
\label{table:specific-value-change}
\end{table*}

In combination with Figure \ref{fig:Changes-in-RPN-index} and Table \ref{table:specific-value-change}, it can be seen that the values of $DR_3$, $DR_5$, and $DR_6$ no longer change significantly, while the values of $DR_1$, $DR_2$ and $DR_4$ form a finite cycle. The three design risks $DR_1$, $DR_2$, and $DR_4$ are impacted by two or more other design risks with varying time delays, resulting in dynamic alterations in the risk index during simulation.

We take the average of the three cycle values of the design risks $DR_1$, $DR_2$, and $DR_4$ as the final value of the risk index after the FCM reaches a stable state (40 minutes). In this way, we obtain the final values $DRPN_i$ of all nodes after the FCM iteration:
\begin{equation*}
DRPN_1:0.7928\ \ \ \ DRPN_2:0.8438\ \ \ \ DRPN_3:0.7815
\end{equation*}
\begin{equation*}
DRPN_4:0.8120\ \ \ \ DRPN_5:0.7020\ \ \ \ DRPN_6:0.7659
\end{equation*}

In addition to the overall iteration, an independent analysis of the design risks is also required to comprehensively evaluate the impact of each design risk on the other design risks. A total of six FCM iterations are performed, only one node is activated each time, and the values of the other nodes are all set to 0.

\begin{equation*}
C_i=\left\{0,\cdots,RPN_i,\cdots,0\right\},(i=1,…,6)
\end{equation*}

When the iteration converges, the DRPN value of each design risk is shown in Table \ref{table:independent-analysis}. All data in bold indicates the largest number in each row, which is the corresponding design risk that is most affected.

\begin{table*}[htb]
\caption{Results of independent analysis of design risks.}
\centering
\begin{tabular}{lllllll} \hline
      & $DR_1$   & $DR_2$  & $DR_3$   & $DR_4$   & $DR_5$   & $DR_6$   \\\hline
Init  & -0.1118 & 0      & 0       & 0       & 0       & 0       \\
Final & 0.7977  & \textbf{0.8549} & 0.7817  & 0.8149  & 0.7001  & 0.7625  \\\hline
Init  & 0       & 0.0417 & 0       & 0       & 0       & 0       \\ 
Final & 0.7980  & 0.8156 & 0.7817  & \textbf{0.8173}  & 0.7012  & 0.7628  \\\hline
Init  & 0       & 0      & -1.3390 & 0       & 0       & 0       \\ 
Final & 0.7976  & \textbf{0.8548} & 0.7817  & 0.8211  & 0.7004  & 0.7619  \\ \hline
Init  & 0       & 0      & 0       & -1.5745 & 0       & 0       \\ 
Final & 0.7961  & \textbf{0.8532} & 0.7813  & 0.8212  & 0.6986  & 0.7570  \\ \hline
Init  & 0       & 0      & 0       & 0       & -1.3381 & 0       \\ 
Final & 0.7963  & \textbf{0.8491} & 0.7817  & 0.8214  & 0.6904  & 0.7555  \\ \hline
Init  & 0       & 0      & 0       & 0       & 0       & -0.8696 \\ 
Final & 0.7979  & \textbf{0.8505} & 0.7817  & 0.8215  & 0.7002  & 0.7574 \\  \hline
\end{tabular}
\label{table:independent-analysis}
\end{table*}

\subsection{Decision making on design risk}
Synthesize the results of Sections 6.1-6.2 to obtain the final risk assessment table as shown in Table \ref{table:final-result}.

\begin{table*}[htb]
\caption{design risk analysis table.}
\centering
\begin{tabular}{clccccc} \hline
No.   & \multicolumn{1}{c}{Design risk Name}       & RPN     & DRPN   & Most Impact DR \\ \hline
$DR_1$ & Inlet Valve failure                  & -0.1118 & 0.7928 & $DR_2$          \\
$DR_2$ & Piston failure                 & 0.0417  & 0.8438 & $DR_4$          \\
$DR_3$ & Cylinder head crackin       & -1.339 & 0.7815 & $DR_2$          \\
$DR_4$ & Fuel injector jam              & -1.5745 & 0.8120 & $DR_2$          \\
$DR_5$ & Big end bearing failure & -1.3381 & 0.702  & $DR_2$          \\
$DR_6$ & Camshaft failure                & -0.8696 & 0.7659 & $DR_2$   \\  \hline
\end{tabular}
\label{table:final-result}
\end{table*}

It can be seen that the initial risk index ($RPN_i$) order for each design risk is as follows:
\begin{equation*}
    DR_2>DR_1>DR_6>DR_5>DR_3>DR_4
\end{equation*}

The initial risk values of $DR_2$ and $DR_1$ are very high, so it is essential to take into account the enhancement of the durability and dependability of the ``Piston'' and ``Valve'' components when producing them.

When considering the risk of system dynamic operation, the order of risk index ($DRPN_i$) is as follows:
\begin{equation*}
    DR_2>DR_4>DR_1>DR_3>DR_6>DR_5
\end{equation*}

It is evident that the ${DR}_4$ risk value has risen drastically, suggesting that it has a higher risk level when the system is in operation. Consequently, it is essential to concentrate on the emergence of the $DR_4:$``Fuel injector jam'' at the system level and to carry out numerous system simulations and optimizations during the system design process to reduce the likelihood of a ``Fuel injector jam'' occurring.

From the last column of Table \ref{table:final-result}, it is evident that the five design risks ${DR}_1$, ${DR}_3$, ${DR}_4$, ${DR}_5$, and ${DR}_6$ all have a greater influence on the design risk ${DR}_2$, and ${DR}_2$ has the most significant effect on ${DR}_4$. Consequently, when constructing the system, it is essential to focus on the reliability of the physical and energy flow of the piston.

In conclusion, when designing and constructing this diesel engine, engineers should pay particular attention to the three risks of the systems $DR_1$, $DR_2$, and $DR_4$, and take steps to address any hidden dangers. Additionally, valves, pistons, and fuel injectors are key systems of the diesel engine.

\subsection{Comparative Analysis}
\subsubsection{Compared with traditional method}
FMEA is a conventional method for risk analysis. Like the Token-FCM analysis approach described in this article, it serves as a qualitative tool to assist designers in pinpointing key failures and adopting preventive strategies early. Consequently, the identical diesel engine example is employed to assess risk using FMEA and is then compared with the method introduced in this article.

Initially, we implement the primary phase of the FMEA analysis, which involves designing the collection of risk data.
As stated in section 6.1, diesel engines exhibit six design risks: ``$DR_1$: inlet valve failure'', ``$DR_2$: piston failure'', ``$DR_3$: cylinder head cracking'', ``$DR_4$: fuel injector jam'', ``$DR_5$: big end bearing failure'', and ``$DR_6$: camshaft failure''. Table 5 displays the O, S, and D metrics, representing the occurrence (O), severity (S), and detection difficulty (D) of the identified faults. Applying defuzzification based on eq.~\ref{eq:plt-dufuzz}, we determine the O, S and D values for each design risk, and these findings are presented in Table\ref{table:FMEA_result}.

Following this, we will assess hazards related to design. In FMEA's approach, risks are assessed from two separate viewpoints. The first viewpoint focuses on examining hazards linked to design risks, whereas the second involves an analysis of risks associated with product functionality. 

When assessing design risk hazards, it is crucial to analyze the fault itself from various perspectives, as well as how the spread of other design risks affects the hazard level. Consequently, the hazard index for each design risk includes two parts: one that evaluates the risk independently ($DRH$) and another that considers the effects of risk propagation ($DRH^*$). These components are calculated using eq.\ref{eq:FMEA_drh} and eq.\ref{eq:FMEA_drh_1}, respectively. Ultimately, the product hazard index ($PH$) is determined using eq.\ref{eq:FMEA_ph}, with all results displayed in Table \ref{table:FMEA_result}.

\begin{subequations}
\begin{equation}
    DRH_{ij}=e^{(O_{ij}+S_{ij}+D_{ij})}
\label{eq:FMEA_drh}
\end{equation}

\begin{equation}
    DRH_{ij}^*=\sum_{k=1}^{n}\sum_{l=1}^{m}{(DRH_{ij}+w_{ijkl}\times DRH_{kl})}
\label{eq:FMEA_drh_1}
\end{equation}

\begin{equation}
    PH_i=\sum_{j=1}^{m}{DRH_{ij}^*}
\label{eq:FMEA_ph}
\end{equation}
\end{subequations}

where $DRH_{ij}$ denotes the design hazard index for the j-th design risk of the i--th function, $e$ is the natural constant, $DRH_{ij}^*$ is the adjusted design hazard index after accounting for risk propagation. The variables $n$ and $m$ indicate the number of functions and the number of risks per function, respectively. The $w_{ijkl}$ represents the impact weight between $DRH_{ij}$ and $DRH_{kl}$, with results sourced from Table \ref{table:impact-relation}. Additionally, $PH_i$ corresponds to the product hazard index of the i-th function.

Referring to the calculation method above, the FMEA results for the diesel engine can be found in the three columns on the right of Table~\ref{table:FMEA_result}. It is evident from the table that the fuel supply function possesses a product hazard index that is markedly higher than that of the transmission function, suggesting the need for a thorough redesign to boost reliability. When looking at individual design risks, piston and intake valve failures have the highest hazard indices, highlighting their significant impact on the entire product. Thus, prioritizing the reliability of the piston and intake valve is crucial in system design. Moreover, taking into account the propagation of risk, the hazard index for a fuel injector jam rises markedly, highlighting that this risk is also a vital factor demanding thorough reliability analysis and design during system development.

When comparing the outcomes of the FMEA analysis with the approach outlined in this paper, as well as Table~\ref{table:FMEA_result} versus Table~\ref{table:final-result}, it becomes evident that both methods yield a similar ranking of risk indicators for design risks considered separately: Piston failure $>$ Inlet valve failure $>$ Camshaft failure $>$ Big end bearing failure $>$ Cylinder head crack $>$ Camshaft failure $>$ Fuel injector jam. The FMEA approach considers the interdependencies of design risks, illustrated in the "Design risk hazard index (Consider fault propagation)" column in Table \ref{table:FMEA_result}, and assesses the cumulative influence of analogous risks within a specific function on the system, as detailed in the product risk index linked to functional division. Nevertheless, this approach concentrates solely on the direct interaction between design risks, overlooking the indirect influences. Specifically, it fails to consider the cumulative effects of risks spreading over time. As seen in the comparative analysis, it undervalues the risk associated with issues like "Fuel injector jam" and "Cylinder head crack".

\begin{small}
 
\begin{table*}[htb]
\centering
\caption{FMEA table of diesel engine system.}
\resizebox{\textwidth}{!}
{
\begin{tabular}{@{}cccccccccc@{}}
\toprule
\multirow{2}{*}{ID} & \multirow{2}{*}{Function}     & \multirow{2}{*}{Design risk name} & \multicolumn{3}{c}{Design risk description} & \multirow{2}{*}{\begin{tabular}[c]{@{}c@{}}Other risk that \\ may cause this risk\end{tabular}}                       & \multirow{2}{*}{\begin{tabular}[c]{@{}c@{}}Design risk \\ hazard index\end{tabular}} & \multirow{2}{*}{\begin{tabular}[c]{@{}c@{}}Design risk hazard index \\ (Consider fault propagation )\end{tabular}} & \multirow{2}{*}{\begin{tabular}[c]{@{}c@{}}Product \\ hazard index\end{tabular}} \\ \cmidrule(lr){4-6}
                    &                               &                                   & O             & S            & D            &                                                                                                                       &                                                                                      &                                                                                                                    &                                                                                  \\ \midrule
1                   & \multirow{3}{*}{Fuel supply}  & Inlet Valve failure               & -0.4          & 0.5          & 0.75         & \begin{tabular}[c]{@{}c@{}}Piston failure,0.6\\ Camshaft failure 0.4\end{tabular}                                     & 2.3396                                                                               & 4.1942                                                                                                             & \multirow{3}{*}{11.5159}                                                         \\ \cmidrule(lr){3-9}
2                   &                               & Piston failure                    & 0.75          & 0.9          & -0.6         & \begin{tabular}[c]{@{}c@{}}Inlet Valve failure,0.8\\ Cylinder head cracking 0.4\end{tabular}                          & 2.8577                                                                               & 4.7835                                                                                                             &                                                                                  \\ \cmidrule(lr){3-9}
3                   &                               & Fuel injector jam                 & 0.15          & -1.5         & -1.25        & \begin{tabular}[c]{@{}c@{}}Piston failure,0.8\\ Cylinder head cracking,0.8\\ Big end bearing failure 0.4\end{tabular} & 0.0743                                                                               & 2.5382                                                                                                             &                                                                                  \\ \midrule
4                   & \multirow{3}{*}{Transmission} & Big end bearing failure           & -1.25         & -0.45        & -0.05        & \begin{tabular}[c]{@{}c@{}}Inlet valve failure,0.2\\ Fuel injector jam 0.6\end{tabular}                               & 0.1738                                                                               & 0.6863                                                                                                             & \multirow{3}{*}{1.3055}                                                          \\ \cmidrule(lr){3-9}
5                   &                               & Cylinder head crackin             & 0.1           & -1.55        & -0.55        & Fuel injector jam 0.4                                                                                                 & 0.1353                                                                               & 0.1650                                                                                                             &                                                                                  \\ \cmidrule(lr){3-9}
6                   &                               & Camshaft failure                  & 0.25          & 0.2          & -1.5         & Big end bearing failure 0.6                                                                                           & 0.3499                                                                               & 0.4542                                                                                                             &                                                                                  \\ \bottomrule
\end{tabular}
}
\label{table:FMEA_result}
\end{table*}
    
\end{small}

\subsubsection{Analysis of the impact of time delay}
Table \ref{table:comparative-analysis} shows the final steady-state value of each design risk modeled for the system without considering the time delay and compares the results with the time delay.

\begin{table*}[htb]
\caption{FCM iteration results considering time delay (left) and not considering time delay.}
\centering
\begin{tabular}{ccc} \hline
No.   & Simulation with   time delays & Simulation   without time delays \\   \hline
$DR_1$ & 0.7928                      & 0.8468                         \\
$DR_2$ & \textbf{0.8438 }                     & 0.9266                         \\
$DR_3$ & 0.7815                      & 0.7946                         \\
$DR_4$ & 0.8120                      & \textbf{0.9302}                         \\
$DR_5$ & 0.7020                       & 0.7058                         \\
$DR_6$ & 0.7659                      & 0.7668   \\  \hline
\end{tabular}
\label{table:comparative-analysis}
\end{table*}

It is obvious from Table \ref{table:comparative-analysis} that time delay plays an important role. Due to the influence of time delays, ${DR}_4$: ``Fuel injector jam'' has not become the design risk with the highest risk index. 

Although $DR_4$ is impacted by the three design risks of $DR_2$, $DR_3$ and $DR_5$, the causal-effects of $DR_2 \rightarrow DR_4$ and $DR_5\rightarrow DR_4$ have a delay of 4 and 6 minutes, respectively, while the time delays of $DR_1 \rightarrow DR_2$ and $DR_3\rightarrow DR_2$ are only 2 and 4 minutes, making the final risk index of $DR_4$ : $DRPN_4$ lower than $DRPN_2$, which is only affected by the two design risks.

Without taking into account the time delay, the risk index of $DR_1$ which is affected by two design risks is much lower than $DR_4$ which is affected by three design risks, and the difference between the two indices is nearly 0.08.

However, due to the two causal-effects of $DR_2\rightarrow DR_1$, $DR_6\rightarrow DR_1$, the time delay is only 2 and 4 minutes, the frequency of ${DR}_1$ affected is higher than ${DR}_4$, resulting in the final risk index of design risks ${DR}_1$ and $DR_4$ being very close when considering the actual time delay.

Obviously, the time delay changes the final steady state value of the FCM node. Furthermore, from simulations with systems that include time delays, it can be observed that systems that do not consider time tend to overestimate the impact of these causal-effect relations. This shows that time relations can play an important role in modeling complex qualitative system dynamics (SD).

\section{Conclusions}
\label{sec:conclusion}
Risk assessment is important for modern engineering design. However, existing methods such as FTA, FMEA, Petri nets, and Bayesian networks, can only handle static and/or one-way causal-effect risk relations in designs. For two-way, dynamic causal-effect risk relations, they have limited applicability. To solve this problem, this paper proposes a new method called Token-FCM to model two-way causal-effect risk relations and to simulate relation dynamics. The main features/contributions of this paper are: (1) a token-augmented FCM method to model two-way and time-delayed causal-effect risk relations in engineering designs; (2) a comprehensive Token-FCM initialization method using the combination of fuzzy sets and group decision-making to collectively, quantitatively and unbiasedly characterize multi-expert opinions on design risk; and (3) a systematic design risk assessment method which can consider static risk indices, dynamic risk indices, and the degree of influence of individual design risks. The method's effectiveness has been validated through a real design example of a diesel engine of horizontal directional drilling machines.

A couple of limitations need to be noted here. The most notable one is that adding tokens to FCM may cause a convergence issue in running Token-FCM. That is, some nodes may not be able to reach a stable state before the maximum iteration number. Although we did not observe this issue in all experiments conducted, this situation could in principle happen if the Token-FCM's time delays are not properly assigned; in particular, different time delays can cause nodes to be activated multiple times in one iteration, resulting in a finite cycle of node values that cannot converge to a fixed value. Analyzing the relationship between time delays and Token-FCM convergence behavior and then designing a convergence-guaranteed token mechanism can be very practically beneficial, and are among the risk assessment research studies to be carried out.

Another limitation is that the current Token-FCM initialization method only makes use of multi-expert opinions. The historical data of an engineering design can be used to enhance the initialization effectiveness. In particular, with the development of artificial intelligence (AI), knowledge of the design's risk probability and fault modes may be learned from the historical data and then used to improve Token-FCM's correctness. Combining such AI techniques with Token-FCM is an interesting and promising improvement direction.

\section*{Acknowledgements}
The authors appreciate the support from the National Key Research and Development Program of China (No. 2023YFB3307202), NSF of China (No. 62102355), the ``Pioneer" and ``Leading Goose" R\&D Program of Zhejiang Province (No. 2024C01103), the Natural Science Foundation of Zhejiang Province (No. LQ22F020012), the Zhejiang Provincial Science and Technology Plan Project (No. 2022C01052), and the Fundamental Research Funds for the Central Universities (No. 226-2023-00020).

%
% ...

\bibliographystyle{asmems4}

\bibliography{asme2e}

\begin{thebibliography}{10}

\bibitem{10.1115/1.2901055}
Qiu, Y., Ge, P., and Yim, S.~C., 2008,
\newblock ``Risk-based resource allocation for collaborative system design in a distributed environment,''
\newblock {\em Journal of Mechanical Design, {\bf 130}}(6), 04, p.~061403.

\bibitem{abdo2016uncertainty}
Abdo, H., and Flaus, J.-M., 2016,
\newblock ``Uncertainty quantification in dynamic system risk assessment: a new approach with randomness and fuzzy theory,''
\newblock {\em International Journal of Production Research, {\bf 54}}(19), pp.~5862--5885.

\bibitem{choi1994petri}
Choi, B.~W., KUOJ, W., and Jackman, J., 1994,
\newblock ``Petri net extensions for modelling and validating manufacturing systems,''
\newblock {\em International Journal of Production Research, {\bf 32}}(8), pp.~1819--1835.

\bibitem{neil2012availability}
Neil, M., and Marquez, D., 2012,
\newblock ``Availability modelling of repairable systems using bayesian networks,''
\newblock {\em Engineering Applications of Artificial Intelligence, {\bf 25}}(4), pp.~698--704.

\bibitem{10.1115/1.4005230}
Coatanéa, E., Nonsiri, S., Ritola, T., Tumer, I.~Y., and Jensen, D.~C., 2011,
\newblock ``A framework for building dimensionless behavioral models to aid in function-based failure propagation analysis,''
\newblock {\em Journal of Mechanical Design, {\bf 133}}(12), 12, p.~121001.

\bibitem{nair2020generalised}
Nair, A., Reckien, D., and van Maarseveen, M.~F., 2020,
\newblock ``Generalised fuzzy cognitive maps: Considering the time dynamics between a cause and an effect,''
\newblock {\em Applied Soft Computing, {\bf 92}}, p.~106309.

\bibitem{kosko1986fuzzy}
Kosko, B., 1986,
\newblock ``Fuzzy cognitive maps,''
\newblock {\em International journal of man-machine studies, {\bf 24}}(1), pp.~65--75.

\bibitem{papageorgiou2012review}
Papageorgiou, E.~I., and Salmeron, J.~L., 2012,
\newblock ``A review of fuzzy cognitive maps research during the last decade,''
\newblock {\em IEEE transactions on fuzzy systems, {\bf 21}}(1), pp.~66--79.

\bibitem{felix2019review}
Felix, G., N{\'a}poles, G., Falcon, R., Froelich, W., Vanhoof, K., and Bello, R., 2019,
\newblock ``A review on methods and software for fuzzy cognitive maps,''
\newblock {\em Artificial intelligence review, {\bf 52}}, pp.~1707--1737.

\bibitem{RUIJTERS201529}
Ruijters, E., and Stoelinga, M., 2015,
\newblock ``Fault tree analysis: A survey of the state-of-the-art in modeling, analysis and tools,''
\newblock {\em Computer Science Review, {\bf 15-16}}, pp.~29--62.

\bibitem{HUANG2020106885}
Huang, J., You, J.-X., Liu, H.-C., and Song, M.-S., 2020,
\newblock ``Failure mode and effect analysis improvement: A systematic literature review and future research agenda,''
\newblock {\em Reliability Engineering \& System Safety, {\bf 199}}, p.~106885.

\bibitem{chandra1983machine}
Chandra, M.~J., and Shanthikumar, J., 1983,
\newblock ``On a machine interference problem with several types of machines attended by a single repairman,''
\newblock {\em International Journal of Production Research, {\bf 21}}(4), pp.~529--541.

\bibitem{saydam2013time}
Saydam, D., Bocchini, P., and Frangopol, D.~M., 2013,
\newblock ``Time-dependent risk associated with deterioration of highway bridge networks,''
\newblock {\em Engineering Structures, {\bf 54}}, pp.~221--233.

\bibitem{laperriere2001monte}
Laperriere, L., and Kabore, T., 2001,
\newblock ``Monte carlo simulation of tolerance synthesis equations,''
\newblock {\em International Journal of Production Research, {\bf 39}}(11), pp.~2395--2406.

\bibitem{10.1115/1.4034219}
Zhu, Z., and Du, X., 2016,
\newblock ``Reliability analysis with monte carlo simulation and dependent kriging predictions,''
\newblock {\em Journal of Mechanical Design, {\bf 138}}(12), 09, p.~121403.

\bibitem{khakzad2013quantitative}
Khakzad, N., Khan, F., and Amyotte, P., 2013,
\newblock ``Quantitative risk analysis of offshore drilling operations: A bayesian approach,''
\newblock {\em Safety science, {\bf 57}}, pp.~108--117.

\bibitem{10.1115/1.4032399}
Yodo, N., and Wang, P., 2016,
\newblock ``Resilience modeling and quantification for engineered systems using bayesian networks,''
\newblock {\em Journal of Mechanical Design, {\bf 138}}(3), 01, p.~031404.

\bibitem{WEISS20051127}
Weiss, L., Amon, C., Finger, S., Miller, E., Romero, D., Verdinelli, I., Walker, L., and Campbell, P., 2005,
\newblock ``Bayesian computer-aided experimental design of heterogeneous scaffolds for tissue engineering,''
\newblock {\em Computer-Aided Design, {\bf 37}}(11), pp.~1127--1139.

\bibitem{kabir2019applications}
Kabir, S., and Papadopoulos, Y., 2019,
\newblock ``Applications of bayesian networks and petri nets in safety, reliability, and risk assessments: A review,''
\newblock {\em Safety science, {\bf 115}}, pp.~154--175.

\bibitem{wu2015extended}
Wu, X.-y., and Wu, X.-Y., 2015,
\newblock ``Extended object-oriented petri net model for mission reliability simulation of repairable pms with common cause failures,''
\newblock {\em Reliability Engineering \& System Safety, {\bf 136}}, pp.~109--119.

\bibitem{10.1115/1.3125203}
Xu, Q., and Jiao, J.~R., 2009,
\newblock ``Modeling the design process of product variants with timed colored petri nets,''
\newblock {\em Journal of Mechanical Design, {\bf 131}}(6), 05, p.~061009.

\bibitem{kelly2009bayesian}
Kelly, D.~L., and Smith, C.~L., 2009,
\newblock ``Bayesian inference in probabilistic risk assessment—the current state of the art,''
\newblock {\em Reliability Engineering \& System Safety, {\bf 94}}(2), pp.~628--643.

\bibitem{mura2001markov}
Mura, I., and Bondavalli, A., 2001,
\newblock ``Markov regenerative stochastic petri nets to model and evaluate phased mission systems dependability,''
\newblock {\em IEEE Transactions on Computers, {\bf 50}}(12), pp.~1337--1351.

\bibitem{bai2005software}
Bai, C., Hu, Q., Xie, M., and Ng, S.~H., 2005,
\newblock ``Software failure prediction based on a markov bayesian network model,''
\newblock {\em Journal of Systems and Software, {\bf 74}}(3), pp.~275--282.

\bibitem{cadini2016bayesian}
Cadini, F., and Gioletta, A., 2016,
\newblock ``A bayesian monte carlo-based algorithm for the estimation of small failure probabilities of systems affected by uncertainties,''
\newblock {\em Reliability Engineering \& System Safety, {\bf 153}}, pp.~15--27.

\bibitem{KIM2012947}
Kim, Y.~S., and Kim, K.-Y., 2012,
\newblock ``Dcr-based causal design knowledge evaluation method and system for future cad applications,''
\newblock {\em Computer-Aided Design, {\bf 44}}(10), pp.~947--960.

\bibitem{papageorgiou2013fuzzy}
Papageorgiou, E.~I., 2013,
\newblock {\em Fuzzy cognitive maps for applied sciences and engineering: from fundamentals to extensions and learning algorithms}, Vol.~54
\newblock Springer Science \& Business Media.

\bibitem{han2018hybrid}
Han, Y., and Deng, Y., 2018,
\newblock ``A hybrid intelligent model for assessment of critical success factors in high-risk emergency system,''
\newblock {\em Journal of Ambient Intelligence and Humanized Computing, {\bf 9}}(6), pp.~1933--1953.

\bibitem{lopez2014dynamic}
Lopez, C., and Salmeron, J.~L., 2014,
\newblock ``Dynamic risks modelling in erp maintenance projects with fcm,''
\newblock {\em Information Sciences, {\bf 256}}, pp.~25--45.

\bibitem{lazzerini2011analyzing}
Lazzerini, B., and Mkrtchyan, L., 2011,
\newblock ``Analyzing risk impact factors using extended fuzzy cognitive maps,''
\newblock {\em IEEE Systems Journal, {\bf 5}}(2), pp.~288--297.

\bibitem{dabbagh2019hybrid}
Dabbagh, R., and Yousefi, S., 2019,
\newblock ``A hybrid decision-making approach based on fcm and moora for occupational health and safety risk analysis,''
\newblock {\em Journal of safety research, {\bf 71}}, pp.~111--123.

\bibitem{wang2009risk}
Wang, Y.-M., Chin, K.-S., Poon, G. K.~K., and Yang, J.-B., 2009,
\newblock ``Risk evaluation in failure mode and effects analysis using fuzzy weighted geometric mean,''
\newblock {\em Expert systems with applications, {\bf 36}}(2), pp.~1195--1207.

\bibitem{wang2021decision}
Wang, G., Wu, L., Liu, Y., and Ye, X., 2021,
\newblock ``A decision-making method for complex system design in a heterogeneous language information environment,''
\newblock {\em Journal of Engineering Design, {\bf 32}}(6), pp.~271--299.

\bibitem{pang2016probabilistic}
Pang, Q., Wang, H., and Xu, Z., 2016,
\newblock ``Probabilistic linguistic term sets in multi-attribute group decision making,''
\newblock {\em Information Sciences, {\bf 369}}, pp.~128--143.

\bibitem{montserrat2018consensus}
Montserrat-Adell, J., Agell, N., S{\'a}nchez, M., and Ruiz, F.~J., 2018,
\newblock ``Consensus, dissension and precision in group decision making by means of an algebraic extension of hesitant fuzzy linguistic term sets,''
\newblock {\em Information Fusion, {\bf 42}}, pp.~1--11.

\bibitem{docent2016approximating}
Docent, D., 2016,
\newblock ``Approximating the condition and maximum load of an engine by comparing engine measurements,''
\newblock PhD thesis, Aalto University.

\end{thebibliography}

\end{document}